\documentclass[reprint,aps,]{revtex4-1}

\usepackage{graphicx}
\usepackage{dcolumn}
\usepackage{bm}
\usepackage{amssymb}
\usepackage{amsfonts}
\usepackage{latexsym}
\usepackage{pdfsync}
\usepackage[T1]{fontenc} 
\usepackage{epsf,amsfonts}
\usepackage{epsfig}
\usepackage{amsthm}
\usepackage{amsmath}
\usepackage{color}
\usepackage{slashed}
\usepackage{mathrsfs}
\usepackage{float}
\usepackage{color}
\usepackage{esvect}
\usepackage{mathtools}
\usepackage{bbm}

\newcommand{\VEV}[1]{\left\langle #1\right\rangle}

\newcommand{\p}{\partial}

\newcommand{\tr}{\mathop{\rm tr}\nolimits}

\newcommand{\MeV}{\;\text{MeV}}
\newcommand{\GeV}{\;\text{GeV}}

\begin{document}

\title{Correlations between the deconfining and chiral transitions in holographic QCD}

\author{Ying-Ying Li}
\email{lyying@hnu.edu.cn}
\affiliation{Department of Applied Physics, School of Physics and Electronics, Hunan University, Changsha 410082, China}

\author{Xing-Lin Liu}
\email{liuxlssy@hnu.edu.cn}
\affiliation{Department of Applied Physics, School of Physics and Electronics, Hunan University, Changsha 410082, China}

\author{Xin-Yi Liu}
\email{liu1347632006@hnu.edu.cn}
\affiliation{Department of Applied Physics, School of Physics and Electronics, Hunan University, Changsha 410082, China}

\author{Zhen Fang}
\thanks{zhenfang@hnu.edu.cn}
\affiliation{Department of Applied Physics, School of Physics and Electronics, Hunan University, Changsha 410082, China}


\begin{abstract}
    We consider an improved soft-wall AdS/QCD model coupled to an Einstein-dilaton system, which offers a way to study the deconfining and chiral transitions simultaneously. The correlation between these two transitions has been investigated in detail in the Einstein-dilaton-scalar system with the bulk scalar field representing the vacuum of matters in the flavor sector of the model. We find that the effects of the scaling dimension $\Delta$ of the dual operator of the dilaton manifest in chiral transitions, although the equation of state can all be matched with the two-flavor lattice results for $\Delta=2.5, 3, 3.5$ in the decoupling case of $\beta=0$. In the weak-coupling case with smaller $\beta$, both the equation of state and the chiral transition exhibit a crossover behavior and turn into first-order phase transitions with the increase of $\beta$.
\end{abstract}

\maketitle

\section{Introduction}\label{introduce1}

QCD phase transition is closely related to the evolution of early universe and the heavy ion collisions at the Relativistic Heavy Ion Collider (RHIC) \cite{Aoki:2006we}. The phase structure of strongly interacting matters is an ongoing hot topic that has lasted for several decades. It is well known that the confinement and chiral symmetry breaking are two fundamental features of low-energy QCD which are related to the hadronic phase at low temperatures. However, with the increase of temperature, the QCD matters will go through the deconfining process, during which the hadronic phase turns into a phase of quark-gluon plasma with a number of new degrees of freedom liberated, and the chiral symmetry will finally be restored as well. Thus, it is natural and critical to study the properties of deconfining and chiral transitions and also their interrelations, which are indeed rather challenging because of the non-perturbative nature of low-energy QCD.

The deconfining phase transition is well defined in the heavy quark limit with the Polyakov loop serving as an order parameter, and it should also be reflected in the behaviors of the equation of state such as the pressure and the energy density and so on. While the chiral phase transition is well defined in the chiral limit with the chiral condensate serving as an order parameter. It has been established by lattice QCD that both deconfining and chiral transitions are analytic crossovers at zero chemical potential with physical quark masses \cite{Aoki:2006we,Bazavov:2011nk,Bhattacharya:2014ara}. However, the nature of the two-flavor chiral transition is still unknown in the chiral limit, although it is generally expected to be a second-order one in the $O(4)$ universality class \cite{Pisarski:1983ms,Burger:2011zc,Pelissetto:2013hqa,Bhattacharya:2014ara}. There are usually two possible scenarios for the QCD phase diagram in the quark-mass plane that need to be settled \cite{Philipsen:2016hkv}. Another interesting issue is on the interrelations between the deconfining and chiral transitions \cite{Sakai:2010rp}. For instance, it is uncertain whether these two types of QCD transitions occur simultaneously or not \cite{Borsanyi:2010bp}.

There have been a large amount of works concentrating on the issue of QCD phase transition, including lattice QCD \cite{Laermann:2003cv,Fukushima:2013rx}, Dyson-Schwinger equations \cite{Fischer:2009wc,Braun:2009gm,Qin:2010nq}, functional renormalization groups \cite{Braun:2009gm}, chiral perturbation theory \cite{Son:2000xc} and so on. Many effective models such as the generalized Nambu Jona-Lasinio model \cite{Ratti:2005jh} and the quark-meson model \cite{Schaefer:2007pw} were also constructed to tackle the relevant problems. In recent decades, the holographic approach, based on the anti-de Sitter/conformal field theory (AdS/CFT) correspondence \cite{Maldacena:1997re,Gubser:1998bc,Witten:1998qj}, has become a powerful tool in the study of nonperturbative QCD. Since AdS/CFT originates from string theory, it would be desirable that the holographic dual of QCD can be constructed from the string-theoretic side \cite{Kruczenski:2003uq,Sakai:2004cn,Sakai:2005yt}. However, this top-down approach is usually unable to provide a good description for the low-energy hadron properties. Hence, most of the holographic studies have adopted a bottom-up approach which is based on the fundamental features of low-energy QCD \cite{deTeramond:2005su,DaRold:2005mxj,Erlich:2005qh,Karch:2006pv,Csaki:2006ji,Cherman:2008eh,Fujita:2009wc,Fujita:2009ca,Colangelo:2009ra,Colangelo:2011sr,Li:2012ay,Li:2013oda,Shuryak:2004cy,Brodsky:2014yha,Tannenbaum:2006ch,Policastro:2001yc,Cai:2009zv,Cai:2008ph,Sin:2004yx,Shuryak:2005ia,Nastase:2005rp,Nakamura:2006ih,Sin:2006pv,Janik:2005zt, Herzog:2006gh,Gursoy:2007cb,Gursoy:2007er,Li:2014hja,Li:2014dsa,Fang:2016cnt,Evans:2016jzo,Mamo:2016xco,Dudal:2016joz,Dudal:2018rki,Ballon-Bayona:2017dvv,Chen:2018msc}. The well-known bottom-up AdS/QCD models are the hard-wall and soft-wall models \cite{DaRold:2005mxj,Erlich:2005qh,Karch:2006pv}. A wide range of low-energy phenomenons have been investigated in the framework of bottom-up AdS/QCD, such as the hadron spectrum \cite{Gherghetta:2009ac,Kelley:2010mu,Sui:2009xe,Sui:2010ay,Cui:2013xva,Cui:2014oba,Fang:2016uer,Fang:2016dqm}, the thermodynamics and particularly the phase structure of QCD \cite{Herzog:2006ra,Gubser:2008yx,Gubser:2008ny}.

It has been shown that the QCD equation of state and the deconfining transition at zero chemical potential can be well described by the Einstein-dilaton system with a proper dilaton potential \cite{Gubser:2008yx,Gubser:2008ny,Noronha:2009ud,Finazzo:2013efa,Finazzo:2014zga,Andreev:2009zk,Yaresko:2013tia,Yaresko:2015ysa,Colangelo:2010pe,Li:2011hp,He:2013qq,Yang:2014bqa,Fang:2015ytf,Rougemont:2017tlu,Li:2017ple,Zollner:2018uep,ChenXun:2019zjc}, while the chiral transition behaviors can be properly characterized in the framework of soft-wall AdS/QCD models \cite{Chelabi:2015gpc,Chelabi:2015cwn,Fang:2016nfj,Li:2016smq,Bartz:2017jku,Bartz:2016ufc,Fang:2018vkp,Fang:2018axm,Fang:2019lmd}. This provides an opportunity for us to study the possible interplay between the deconfining and chiral transitions by combining the Einstein-dilaton system with the soft-wall models. In order to give a complete description for QCD phase transition in the holographic framework, we have proposed an improved soft-wall AdS/QCD model with the background fields solved from an Einstein-dilaton system, which is able to characterize the deconfining and chiral transitions qualitatively in the two-flavor case with nonzero quark masses \cite{Fang:2019lsz}. However, the bulk background in this model is independent of the flavor sector, and thus the issue on the correlation of these two transitions cannot be addressed in that work.

To investigate the interrelations between the deconfining and chiral transitions, we need to consider the Einstein-dilaton-scalar system with the background fields coupled to the vacuum of matter fields, which is indeed not that easy to solve in numerics. In this work, we would like to give a detailed analysis on this coupled system, in which the back-reaction of the vacuum of matters to the bulk background will be fully addressed. In addition, we will also consider the effects of the scaling dimension $\Delta$ of the dual operator of the dilaton on QCD phase transition. Though there have been many discussions on the physical relevance of the two-dimension operator related to $A^{2}$ \cite{Gubarev:2000eu}, it is generally believed that the most natural candidate for the dual of the dilaton is the local gauge-invariant gluon operator $\tr F^2_{\mu\nu}$, the dimension of which should depend on the energy scale that has been taken. Here we shall only concern ourselves with the scaling dimension $\Delta$, rather than the specific dual operator of the dilaton. Previous studies indicate that we cannot distinguish different values of $\Delta$ in the Breitenlohner-Freedman (BF) bound through thermodynamic properties such as the equation of state obtained from the Einstein-dilaton system, which can be attributed to the redundant degrees of freedom embodied in the paramters of the dilaton potential \cite{Gubser:2008ny,Noronha:2009ud,Finazzo:2013efa,Finazzo:2014zga}. However, this situation would be changed once we take the flavor sector into account and consider the chiral dynamics of the Einstein-dilaton-scalar system, which naturally sets a physical energy scale related to the low-energy hadron physics.

It should be remarked that the interplays between the gluon and chiral dynamics have also been analyzed in the holographic models in the Veneziano limit (V-QCD) \cite{Jarvinen:2011qe,Alho:2012mh}. Unlike the Einstein-dilaton-scalar system based on an improved soft-wall model with a specific Einstein-dilaton background system, the V-QCD models have combined the improved holographic QCD for pure gluon dynamics \cite{Gursoy:2008bu} with a tachyon Dirac-Born-Infeld action which controls the dynamics of chiral symmetry breaking. The thermodynamics and the chiral transition have been considered in this framework at finite temperature and density \cite{Alho:2013hsa}, with a rather different phase structure from that will be displayed in this work.

The paper is organized as follows. In Sec. \ref{sec-model}, we give a brief outline of the improved soft-wall AdS/QCD model coupled to an Einstein-dilaton system, and then we focus on the Einstein-dilaton-scalar system that will be mainly addressed in this work. In Sec. \ref{sec-eom-bc}, we derive the equation of motion (EOM) of the bulk fields from the coupled system and specify the boundary conditions for the cases of $\Delta=2.5, 3, 3.5$. In Sec. \ref{eos-pt}, we investigate the behaviors of the equation of state and chiral transition for each case of $\Delta$ with different values of the coupling constant $\beta$. In Sec. \ref{sec-conc}, we conclude our work with a few discussions.

\section{The improved soft-wall model coupled with an Einstein-dilaton system}\label{sec-model}

\subsection{Model action}


We consider an improved soft-wall AdS/QCD model coupled to an Einstein-dilaton system which determines the profiles of the background fields \cite{Fang:2019lsz}. The bulk background is dual to the pure Yang-Mills sector of QCD which incorporates informations of the gluon dynamics, while the flavor sector of the improved soft-wall model chracterizes the low-energy hadron properties. The metric ansatz for the background geometry can be written in the string frame as
\begin{align}\label{stringmetric}
    ds^2 & = \frac{L^2 e^{2 A_S(z)}}{z^2} \left(-f(z)dt^2 + dx^i dx^i +\frac{dz^2}{f(z)}\right)
\end{align}
with an asymptotic AdS structure in the ultraviolet (UV) region ($z\to 0$), and the AdS radius will be set to $L=1$ for simplicity.

The bulk action of the whole system can be decomposed into two parts:
\begin{align}\label{act-grav-2scalar1}
    S=S_G+S_M ,
\end{align}
where the background sector $S_G$ is just the action of the Einstein-dilaton system:
\begin{align}\label{act-grav1}
    S_G & = \frac{1}{2\kappa_5^2}\int d^5x\sqrt{-g}e^{-2 \phi }\left[R +4(\partial\phi)^2 -V(\phi)\right] ,
\end{align}
where $\kappa_5^2=8\pi G_5$, and a nontrivial dilaton potential $V(\phi)$ needs to be specified later. An appropriate form of $V(\phi)$ will generate the relevant deformations of the dual conformal field theory so as to reproduce the expected thermodynamics of QCD. The flavor sector $S_M$ represents the action of the improved soft-wall AdS/QCD model which can be written as
\begin{align}\label{act-flav1}
    S_M & = -\kappa\int d^5x\sqrt{-g}e^{-\phi}\mathrm{Tr}\Big\{|DX|^2 +V_X(X,\phi)      \nonumber \\
        & \quad +\frac{1}{4g_5^2}(F_L^2+F_R^2)\Big\} ,
\end{align}
where $D^MX=\p^MX -iA_L^MX+i X A_R^M$ and $F_{L,R}^{MN}=\partial^MA_{L,R}^N-\partial^NA_{L,R}^M-i[A_{L,R}^M,A_{L,R}^N]$ with the gauge fields $A_{L,R}^M$ in the adjoint representation of $\mathrm{SU}(2)_{L,R}$, and the potential of the bulk scalar field $X$ takes the form
\begin{align}\label{VX2}
    V_X(X,\phi) =m_5^2|X|^{2} -\lambda_1\phi |X|^{2} +\lambda_2|X|^{4},
\end{align}
where a cubic coupling term between the bulk scalar field $X$ and the dilaton $\phi$ has been added in order to realize the correct behaviors of chiral transition in this improved soft-wall AdS/QCD model \cite{Fang:2019lsz}. The mass squared of the bulk scalar field $X$ is determined by the mass-dimension relation $m_5^2L^2 =\Delta_X(\Delta_X-4)$ with $\Delta_X=3$ being the scaling dimension of the dual operator $\bar{q}_Rq_L$ of the scalar field $X$ in the boundary \cite{Erlich:2005qh}.

\subsection{The Einstein-dilaton-scalar system}

According to Ref. \cite{Erlich:2005qh}, the vacuum expectation value (VEV) of the bulk scalar field $X$ can be written as $\langle X\rangle=\frac{\chi(z)}{2}I_2$ with $I_2$ denoting the $2\times2$ unit matrix, and the chiral condensate $\sigma$ is embodied in the UV expansion of the scalar VEV $\langle X\rangle$. Hence, in order to investigate the properties of chiral transition, we only need to consider the vacuum part of matter fields represented by $\langle X\rangle$ in the bulk action (\ref{act-flav1}) and neglect the vacuum fluctuations corresponding to the meson fields. The bulk action (\ref{act-grav-2scalar1}) will then be reduced to
\begin{align}\label{Ein-two-scal-str1}
    S & =S_G+S_{\chi}      \nonumber                                                                          \\
      & =\frac{1}{2\kappa_5^2}\int d^5x\sqrt{-g}e^{-2\phi}\Big[R +4(\partial\phi)^2 -V(\phi)        \nonumber \\
      & \quad -\beta e^{\phi}\Big(\frac{1}{2}(\partial\chi)^2 +V(\chi,\phi)\Big)\Big] ,
\end{align}
where $\beta=16\pi G_5\kappa$, and the potential term of the scalar VEV $\chi$ takes the form
\begin{align}\label{Vchi1}
    V(\chi,\phi) & =\mathrm{Tr}\,V_X(\VEV{X},\phi)        \nonumber                          \\
                 & =\frac{1}{2}(m_5^2-\lambda_1\phi)\chi^{2} +\frac{\lambda_2}{8} \chi^{4} .
\end{align}
The reduced action (\ref{Ein-two-scal-str1}) is just the action of the so-called Einstein-dilaton-scalar system which incorporates both the informations of deconfinement and that of chiral transition. Note that the parameter $\beta$ quantifies the coupling strength between the scalar VEV $\chi$ and the bulk background, which signifies the entanglement between the chiral and deconfining transitions.

For convenience, we perform the calculation in the Einstein frame with the following metric ansatz:
\begin{align}\label{einst-metric}
    ds^2 & = \frac{L^2 e^{2 A_E(z)}}{z^2} \left(-f(z)dt^2 + dx^i dx^i +\frac{dz^2}{f(z)}\right) ,
\end{align}
where the warp factor $A_E$ is related to $A_S$ by $A_E=A_S -\frac{2}{3}\phi$. The action (\ref{Ein-two-scal-str1}) in the string frame can then be transformed into the Einstein frame with the form
\begin{align}\label{Ein-two-scal-ef1}
    S & =S_G+S_{\chi}      \nonumber                                                                                \\
      & =\frac{1}{2\kappa_5^2}\int d^5x\sqrt{-g_E}\Big[R_E -\frac{4}{3}(\partial\phi)^2 -V_E(\phi)        \nonumber \\
      & \quad -\beta e^{\phi}\Big(\frac{1}{2}(\partial\chi)^2 +V_E(\chi,\phi)\Big)\Big],
\end{align}
where
\begin{align}\label{Vphi-Vchi}
    V_E(\phi)= e^{\frac{4\phi}{3}}V(\phi), \quad\,\,\,  V_E(\chi,\phi) = e^{\frac{4\phi}{3}}V(\chi,\phi).
\end{align}

\subsection{The dilaton potential}

Now we specify the form of the dilaton potential $V_E(\phi)$ which is critical to realize the thermodynamic properties of QCD  and particularly the equation of state addressed in this work. With a rescaling of the dilaton $\phi_c= \sqrt{8/3}\,\phi$, the background sector of the action (\ref{Ein-two-scal-ef1}) can be recast into the canonical form
\begin{align}\label{cano-gra-act1}
    S_G & =\frac{1}{2\kappa_5^2}\int d^5x\sqrt{-g_E}\left[R_E -\frac{1}{2}(\partial\phi_c)^2 -V_c(\phi_c)\right]
\end{align}
with $V_c(\phi_c) =V_E(\phi)$. The asymptotic AdS structure of the bulk geometry requires the following UV expansion of the dilaton potential $V_c(\phi_c)$:
\begin{align}\label{ein-V-phi-UV}
    V_c(\phi_c\to0) \simeq -\frac{12}{L^2}+\frac{1}{2}m^2\phi_c^2+\mathcal{O}(\phi_c^4).
\end{align}
Following Ref. \cite{Gubser:2008ny}, we will choose a dilaton potential with exponential form in the infrared (IR) region, i.e., $V_c(\phi_c) \sim V_0e^{\gamma\phi_c}$ with $V_0<0$ and $\gamma>0$, which corresponds to the Chamblin-Reall solution \cite{Chamblin:1999ya}. It has been shown that the adiabatic generalization of such a solution is able to mimick the equation of state from lattice QCD.

According to AdS/CFT, the scaling dimension $\Delta$ of the dual operator of $\phi_c$ is connected with the bulk mass of $\phi_c$ through the mass-dimension relation $m^2 L^2= \Delta(\Delta-4)$ with $\Delta$ constrained in the BF bound \cite{Breitenlohner:1982bm}. The special case of $\Delta=3$ has been considered in Ref. \cite{Fang:2019lsz}, where it was shown that the Einstein-dilaton system with a proper dilaton potential can be used to reproduce the QCD equation of state and other thermodynamic properties \cite{Finazzo:2013efa,Finazzo:2014zga}. In addition, the right behaviors of chiral transition can also be realized qualitatively in the improved soft-wall AdS/QCD model with the action (\ref{act-flav1}) under the background solved from the Einstein-dilaton system. Indeed, similar results for the equation of state can also be obtained by taking other values of $\Delta$ which may be regarded as the dimension of the gluon operator $\tr F^2_{\mu\nu}$ at different energy scales \cite{Gubser:2008yx}. To check this further, we will analyze three cases with $\Delta=2.5, 3, 3.5$ in this work.

In the light of the UV and IR asymptotic forms of $V_c(\phi_c)$, we just adopt the dilaton potential given in Ref. \cite{Gubser:2008yx} with the simpler form
\begin{align}\label{phi-potent1}
    V_c(\phi_c) =\frac{1}{L^2}\left(-12\cosh\gamma\phi_c +b_2\phi_c^2 +b_4\phi_c^4\right)
\end{align}
which has the following UV expansion
\begin{align}\label{phi-potent-uv1}
    V_c(\phi_c\to 0) \simeq \frac{-12}{L^2} +\frac{b_2-6\gamma^2}{L^2}\phi_c^2 +\mathcal{O}(\phi_c^4) ,
\end{align}
in which the parameters $b_2$ and $\gamma$ should be related to each other by
\begin{align}\label{gamma-b2}
    b_2 =6\gamma^2 +\frac{\Delta(\Delta-4)}{2} .
\end{align}
This simpler form of $V_c(\phi_c)$ will be shown to mimick the equation of state from two-flavor lattice QCD quite well for all the cases of $\Delta=2.5, 3, 3.5$ in the decoupling limit of $\beta=0$.

\section{Equation of motion and boundary condition}\label{sec-eom-bc}

\subsection{Equation of motion}\label{sec-eom}

We derive the EOMs for the bulk fields of the Einstein-dilaton-scalar system by the variation of the action (\ref{Ein-two-scal-ef1}) with respect to these fields. The Einstein equation can be obtained as
\begin{align}\label{eins-eq0}
     & R_{MN} -\frac{1}{2}g_{MN}R +\frac{4}{3}\left(\frac{1}{2}g_{MN}\partial_J\phi\,\partial^J\phi -\partial_M\phi\partial_N\phi\right)      \nonumber              \\
     & +\frac{1}{2}g_{MN}V_E(\phi) +\frac{\beta}{2}e^{\phi}\left(\frac{1}{2}g_{MN}\partial_J\chi\,\partial^J\chi -\partial_M\chi\partial_N\chi\right)      \nonumber \\
     & +\frac{\beta}{2}g_{MN}e^{\phi}V_E(\chi,\phi) =0
\end{align}
which contains two independent equations of the form
\begin{align}
    f'' +3A_E'f' -\frac{3}{z}f'                                                        & = 0 ,  \label{fz-eom2} \\
    A_E'' +\frac{2}{z}A_E' -A_E'^2 +\frac{4}{9}\phi'^2 +\frac{\beta}{6}e^{\phi}\chi'^2 & = 0 .  \label{AE-eom2}
\end{align}
The EOMs of the dilaton $\phi$ and the scalar VEV $\chi$ take the form
\begin{align}
    \phi'' +\left(3 A_E' +\frac{f'}{f}-\frac{3}{z}\right)\phi' -\frac{3\beta}{16}e^{\phi}\chi'^2 \qquad                                 & \nonumber                     \\
    -\frac{3e^{2A_E}\partial_{\phi}V_E(\phi)}{8z^2f} -\frac{3\beta e^{2A_E}\partial_{\phi}\left(e^{\phi}V_E(\chi,\phi)\right)}{8 z^2 f} & =0 ,  \label{dilaton-eom2}    \\
    \chi''+\left(3A_E' +\phi' +\frac{f'}{f}-\frac{3}{z}\right)\chi'                                                                     & \nonumber                     \\
    -\frac{e^{2A_E}\,\partial_{\chi}V_E(\chi,\phi)}{z^2 f}                                                                              & =0 .   \label{scalarvev-eom2}
\end{align}

The profiles of the background fields $A_E$, $f$, $\phi$ and the scalar VEV $\chi$ can be obtained by solving the coupled Eqs. (\ref{fz-eom2}) - (\ref{scalarvev-eom2}) numerically with appropriate boundary conditions, which is yet not an easy work. To further simplify the computation, we may substitute Eq. (\ref{fz-eom2}) with a first-order differential equation
\begin{align}\label{fz-eom3}
    f' +4f_4 e^{-3A_E}z^3 = 0 ,
\end{align}
where $f_4$ is an integration constant.

\subsection{The boundary conditions}\label{sec-bc}

We specify the boundary conditions that will be used to obtain reasonable solutions of the Einstein-dilaton-scalar system. At finite temperature, the bulk geometry of the form (\ref{einst-metric}) is a black hole solution with an event horizon $z_h$ and approaches AdS$_5$ asymptotically in the UV limit $z\to 0$, which leads to the following boundary conditions for $f(z)$:
\begin{align}\label{bc-f-z}
    f(0) =1, \qquad   f(z_h)=0.
\end{align}

The other boundary conditions will be taken from the UV expansions of the dilaton $\phi$ and the scalar VEV $\chi$, which can be obtained from the asymptotic analysis of Eqs. (\ref{fz-eom2}) - (\ref{scalarvev-eom2}). Note that the UV forms of the bulk fields depend on the scaling dimension $\Delta$ of the dual operator of $\phi(z)$. For the case of $\Delta=3$, the UV expansions of the bulk fields at $z\to0$ take the forms
\begin{align}
    f(z)    & = 1 -f_4 z^4 +\cdots ,     \label{f-uv3}                                                                                                  \\
    A_E(z)  & = -\frac{1}{108}\left(3\beta m_q^2\zeta^2 +8 p_1^2\right)z^2      \nonumber                                                               \\
            & \quad -\frac{1}{24}\beta p_1 m_q^2\zeta^2 (2\lambda_1+11) z^3 +\cdots ,    \label{AE-uv3}                                                 \\
    \phi(z) & = p_1 z +\frac{3}{16} \beta m_q^2\zeta ^2 (\lambda_1+6)z^2 +p_3 z^3       \nonumber                                                       \\
            & \quad -\left[\frac{1}{48} \beta p_1 m_q^2\zeta ^2 \left(9\lambda_1^2 +111 \lambda_1 +286\right) \right.      \nonumber                    \\
            & \quad \left. -\frac{4}{9} p_1^3 \left(12 b_4-6 \gamma ^4+1\right)\right] z^3\ln z +\cdots ,
    \label{phi-uv3}                                                                                                                                     \\
    \chi(z) & =m_q\zeta z +p_1 m_q\zeta(\lambda_1 +5) z^2 +\frac{\sigma}{\zeta} z^3       \nonumber                                                     \\
            & \quad  -\left[\frac{1}{96} m_q^3\zeta^3 \left(\beta  \left(9\lambda_1^2+108 \lambda_1 +308\right)-24 \lambda_2\right) \right.   \nonumber \\  &\quad \left. +\frac{1}{18}p_1^2m_q\zeta \left(9\lambda_1^2 +111 \lambda_1+286\right) \right]z^3\ln z +\cdots ,   \label{chi-uv3}
\end{align}
while for the case of $\Delta =2.5$, we have
\begin{align}
    f(z)    & = 1 -f_4 z^4 +\cdots ,     \label{f-uv52}                                                                       \\
    A_E(z)  & = -\frac{1}{36}\beta m_q^2\zeta^2 z^2 -\frac{1}{12}p_1^2 z^3 +\cdots ,    \label{AE-uv52}                       \\
    \phi(z) & = p_1 z^{3/2} +\frac{3}{4}\beta m_q^2\zeta^2 (\lambda_1 +6) z^2 +p_3 z^{5/2} +\cdots ,
    \label{phi-uv52}                                                                                                          \\
    \chi(z) & = m_q\zeta z +\frac{2}{3}p_1 m_q\zeta (2\lambda_1 +11) z^{5/2} +\frac{\sigma}{\zeta} z^3       \nonumber        \\
            & \quad  -\frac{1}{24} m_q^3\zeta^3 \left[\beta\left(9\lambda_1^2 +108\lambda_1 +320\right) \right.     \nonumber \\  &\quad \left. -6 \lambda_2\right] z^3\ln z +\cdots,     \label{chi-uv52}
\end{align}
where $m_q$ denotes the quark mass, $\sigma$ denotes the chiral condensate and $\zeta= \frac{\sqrt{N_c}}{2\pi}$ is a normalization constant \cite{Cherman:2008eh}. The coefficient $f_4$ is connected with the event horizon $z_h$ which is further related to the temperature $T$. The UV asymptotic forms of $\phi(z)$ contain two other independent coefficients $p_1$ and $p_3$ which should also be specified. The values of $m_q$ and $p_1$ prescribe the remaining two boundary conditions for solving Eqs. (\ref{fz-eom2}) - (\ref{scalarvev-eom2}). As for the case of $\Delta =3.5$, the UV forms of the bulk fields are not presented here due to the lengthy expressions, and only the case of $m_q =0$ will be addressed for $\Delta =3.5$ since the chiral condensate cannot be extracted with enough degree of accuracy in the case of $m_q\neq 0$ by the numerical method used in this work.

In the numerical calculation, we define another two fields in place of the dilaton $\phi$ and the scalar VEV $\chi$ by
\begin{align}\label{phichi-phitchit}
     & \tilde{\phi} \equiv \frac{z_h}{z}\phi, \quad  \tilde{\chi} \equiv \frac{z_h}{z}\chi  \qquad \text{for $\Delta=3$,}     \nonumber                             \\
     & \tilde{\phi} \equiv \left(\frac{z_h}{z}\right)^{\frac{3}{2}}\phi, \quad  \tilde{\chi} \equiv \frac{z_h}{z}\chi  \qquad \text{for $\Delta=2.5$,}              \\
     & \tilde{\phi} \equiv \left(\frac{z_h}{z}\right)^{\frac{1}{2}}\phi, \quad  \tilde{\chi} \equiv \frac{z_h}{z}\chi  \qquad \text{for $\Delta=3.5$,}    \nonumber
\end{align}
the values of which at $z=0$ will then be taken as the UV boundary conditions:
\begin{align}\label{bc-phitchit}
     & \tilde{\phi}(0) =p_1 z_h, \quad  \tilde{\chi}(0) =m_q\zeta z_h  \qquad \text{for $\Delta=3$,}    \nonumber                \\
     & \tilde{\phi}(0) =p_1 z_h^{\frac{3}{2}}, \quad  \tilde{\chi}(0) =m_q\zeta z_h  \qquad \text{for $\Delta=2.5$,}             \\
     & \tilde{\phi}(0) =p_1 z_h^{\frac{1}{2}}, \quad  \tilde{\chi}(0) =m_q\zeta z_h  \qquad \text{for $\Delta=3.5$.}   \nonumber
\end{align}
To simplify the calculation, we also replace the variable $z$ by a new variable $t$ with the relation
\begin{align}\label{varepl-z-t}
    z = z_h \frac{t+1}{2}, \qquad   -1\leq t \leq 1.
\end{align}

\section{Equation of state and phase transition}\label{eos-pt}

Now we consider the equation of state and the phase transition in the Einstein-dilaton-scalar system with the action (\ref{Ein-two-scal-ef1}). Specifically, we will calculate the entropy density, the pressure, the energy density and the trace anomaly, and investigate the behaviors of these thermodynamic observables with respect to temperature, which reflect the properties of deconfinement. For the coupled system, the vacuum of matters represented by the scalar VEV $\chi$ will have a back-reaction to the background fields, and thus has an unignorable influence on QCD thermodynamics. This back-reaction effect of the flavor part on the bulk background will be investigated in detail, along with the properties of chiral transition that is embodied in the scalar VEV $\chi$, which allows us to probe into the issue on the correlations between the deconfining and chiral transitions.

As aforementioned, the bulk geometry is a black hole with an event horizon $z_h$ such that $f(z_h)=0$. According to AdS/CFT, the temperature $T$ of the system is given by the Hawking formula
\begin{align}\label{temperat-T}
    T=\frac{|f'(z_h)|}{4\pi},
\end{align}
and the entropy density $s$ of the system is given by the formula
\begin{align}\label{entropy-S}
    s =\frac{2\pi e^{3A_E(z_h)}}{\kappa_5^2 z_h^3}.
\end{align}
The pressure $p$ of the system can then be obtained from the thermodynamic relation $s=\partial p/\partial T$ with fixed chemical potential:
\begin{align} \label{p-T}
    p =-\int_{\infty}^{z_h} s(\tilde{z}_h)T'(\tilde{z}_h) d\tilde{z}_h ,
\end{align}
through which the energy density $\varepsilon=-p+sT$ and the trace anomaly $\varepsilon-3p$ can also be obtained.

These thermodynamic observables will be computed separately for the cases of $\Delta=2.5, 3, 3.5$. As a remark, one of the reasons to choose three values of $\Delta$ in our case is to check that such an Einstein-dilaton-scalar system can reproduce almost equally well the QCD equation of state and other thermodynamic quantities for different values of $\Delta$ in the BF bound $2<\Delta<4$ as long as the parameters of the dilaton potential $V(\phi)$ are adjusted appropriately. Thus we cannot determine the most proper one of $\Delta$ by only considering the equation of state in the framework of our model and many other ones. However, once the bulk background were fixed by the QCD equation of state, the effect of the scaling dimension $\Delta$ on chiral transition would be shown manifestly, as will be seen below.

\subsection{$\Delta=3$}

We first investigate the case of $\Delta=3$ which has been addressed in Ref. \cite{Fang:2019lsz} without consideration of the back-reaction of the scalar VEV to the background, which just corresponds to the decoupling case of $\beta=0$ in this work. With the boundary conditions (\ref{bc-f-z}) and (\ref{bc-phitchit}), we are able to solve Eqs. (\ref{fz-eom2}) - (\ref{scalarvev-eom2}) numerically to obtain the profiles of the bulk fields, and thereby the equation of state can be computed. It is reasonable to assume that the back-reaction effect will not be large, so that we only consider three cases with the coupling constant $\beta=0, 0.2, 0.4$. We fit the equation of state obtained from the model with the two-flavor lattice results in the decoupling case of $\beta=0$ with $m_q=5\MeV$, as in Ref. \cite{Fang:2019lsz}. The parameters in the dilaton potential (\ref{phi-potent1}) are set to $\gamma=0.55$ and $b_4=-0.125$, and the parameter $p_1$ in the UV form of the dilaton $\phi$ is set to $p_1=0.675 \GeV$. The coupling constants in the scalar potential (\ref{Vchi1}) will be taken as $\lambda_1=-1.2$ and $\lambda_2=1$ throughout the paper. The influences of $\lambda_1$ and $\lambda_2$ on chiral transition behaviors have been investigated in Ref. \cite{Fang:2019lsz}. In addition, all the observables will be computed for both the case of $m_q=0$ and the case of $m_q=5\MeV$.

\begin{figure}
    \centering
    \includegraphics[width=75mm,clip=true,keepaspectratio=true]{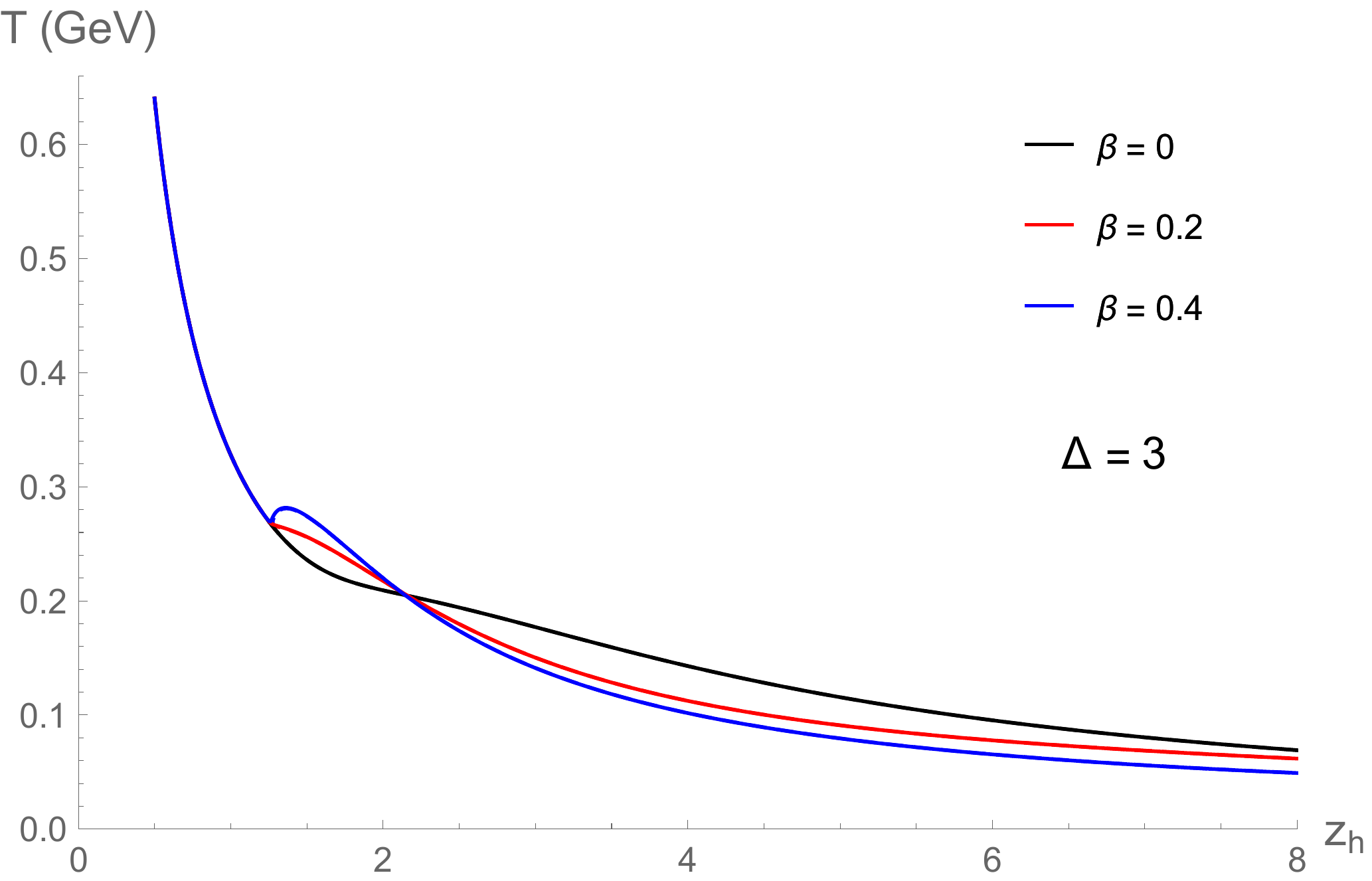}
    \vskip 0.3cm
    \caption{The variations of temperature $T$ with respect to the horizon $z_h$ for $\beta=0, 0.2, 0.4$ in the case of $\Delta=3$.}
    \label{fig-T-zh3}
\end{figure}

The temperature $T$ as a function of the horizon $z_h$ for $\beta=0, 0.2, 0.4$ has been shown in Fig. \ref{fig-T-zh3}, where we can see that $T$ decreases monotonically with the increase of $z_h$ in the decoupling case of $\beta=0$, while this monotonicity changes in some range of $z_h$ when $\beta$ increases to larger values, which, as a result, will change the order of phase transition, as will be shown later.

\begin{figure}
    \begin{center}
        \includegraphics[width=68mm,clip=true,keepaspectratio=true]{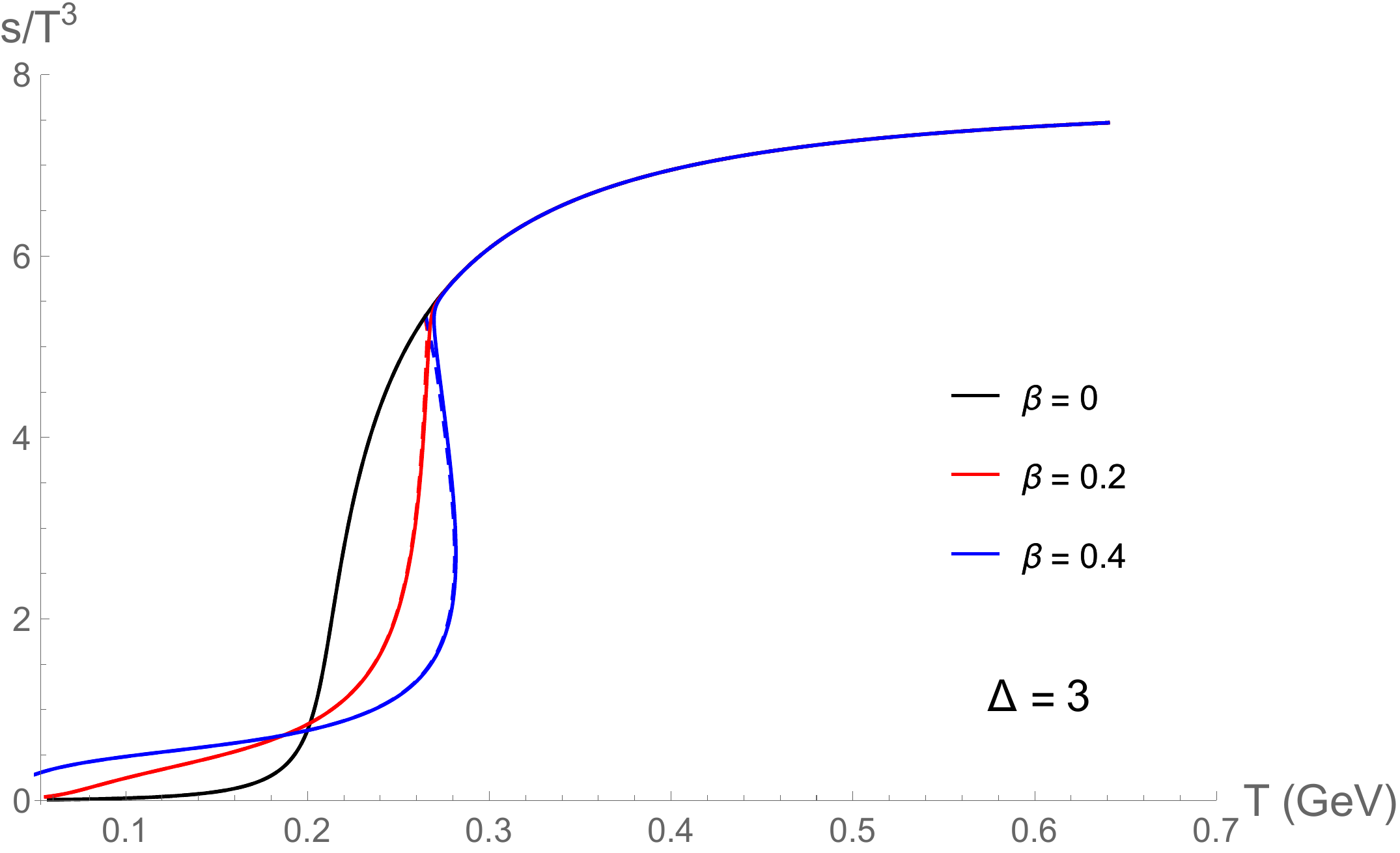}
        \vskip 0.5cm
        \includegraphics[width=68mm,clip=true,keepaspectratio=true]{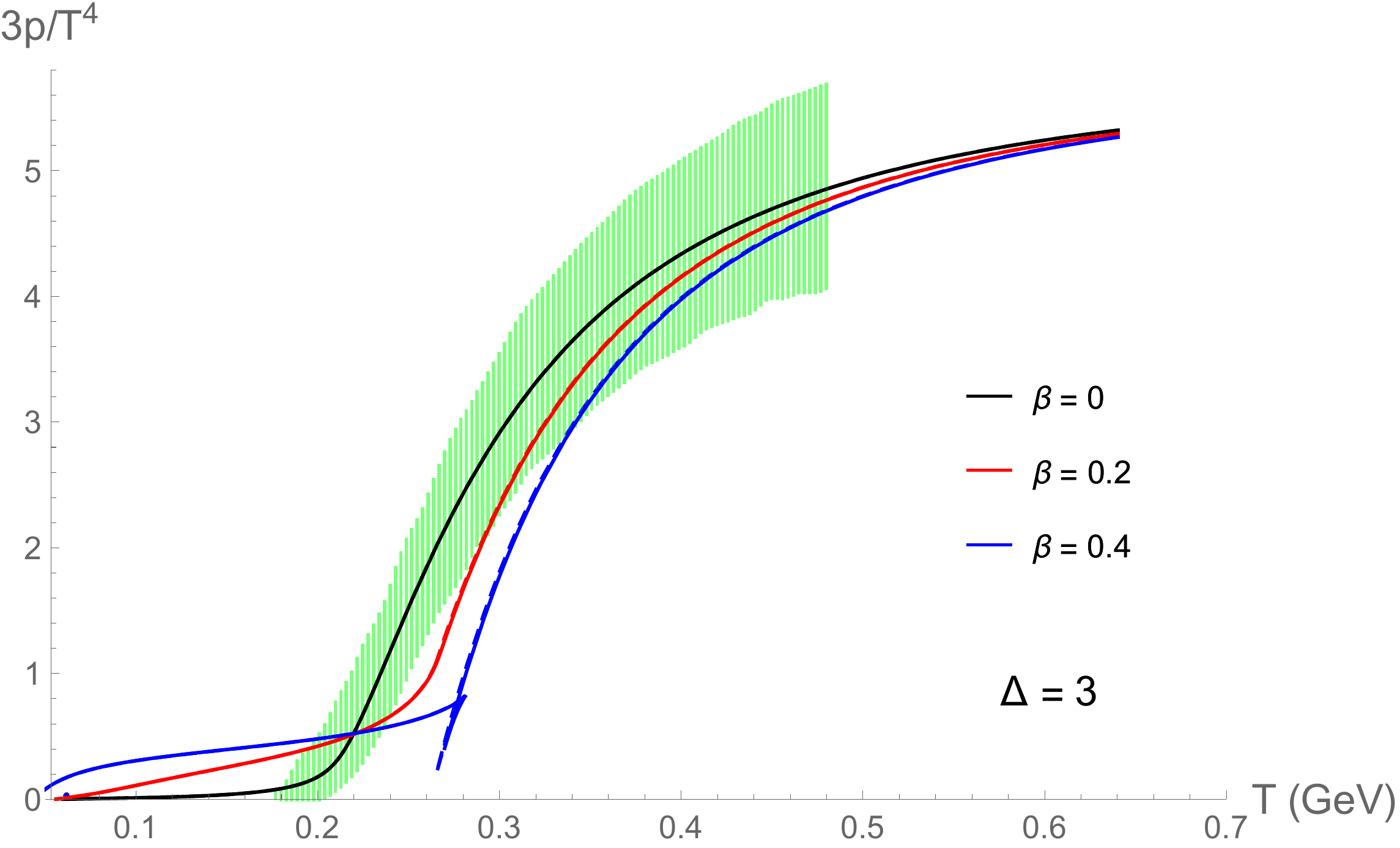} \vskip 0.3cm 
    \end{center}
    \caption{The behaviors of the rescaled entropy density $s/T^3$ (upper panel) and pressure $3p/T^4$ (lower panel) with respect to the temperature $T$ for $\beta=0, 0.2, 0.4$ in the case of $\Delta=3$. The green bands represent the lattice interpolations of two-flavor QCD \cite{Burger:2014xga}. The dashed curves denote the case of $m_q=0$ and the solid ones denote the case of $m_q=5\MeV$.}
    \label{fig-s-p-T3}
\end{figure}

\begin{figure}
    \begin{center}
        \includegraphics[width=68mm,clip=true,keepaspectratio=true]{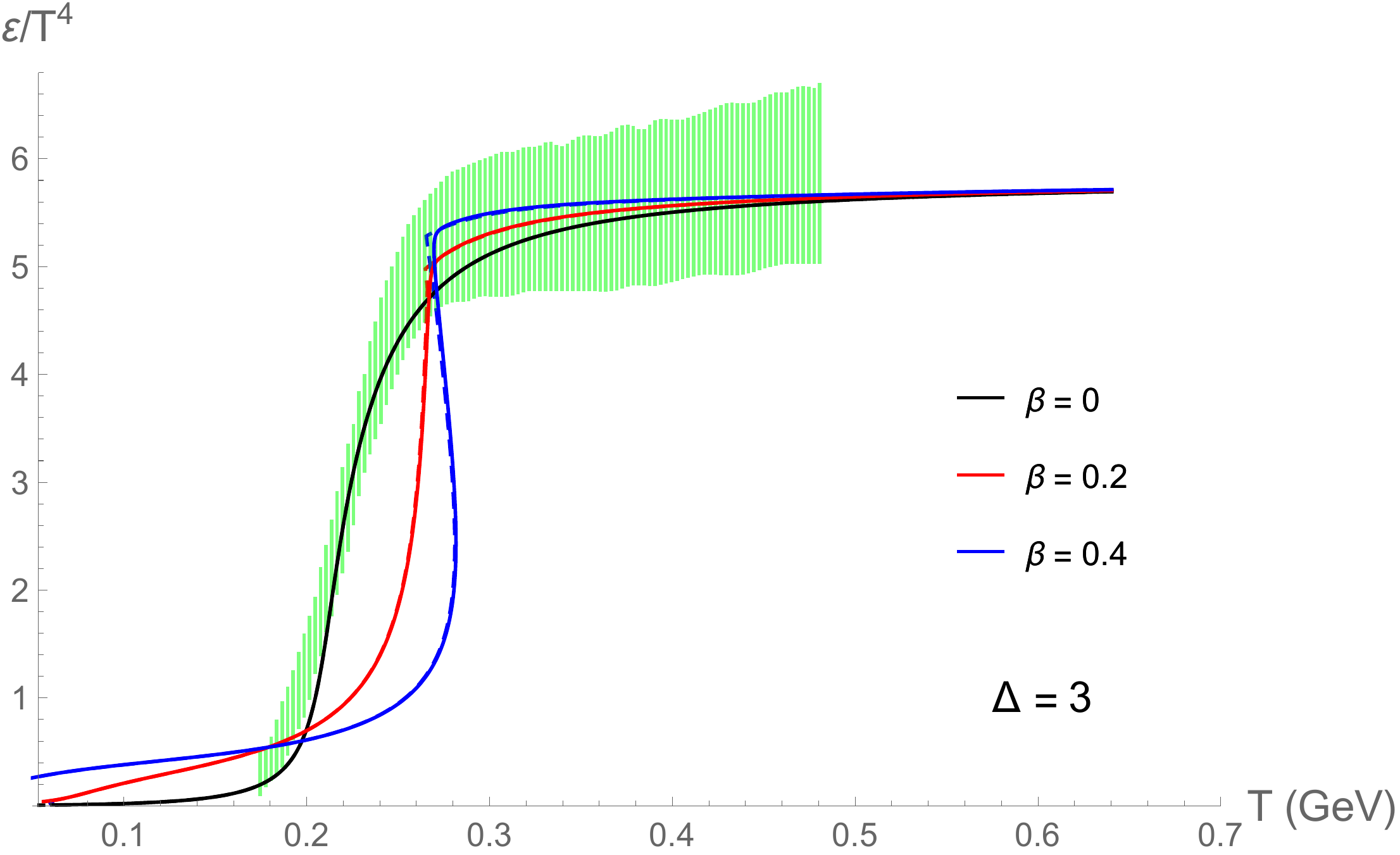}
        \vskip 0.5cm\hspace*{-0.5cm}
        \includegraphics[width=68mm,clip=true,keepaspectratio=true]{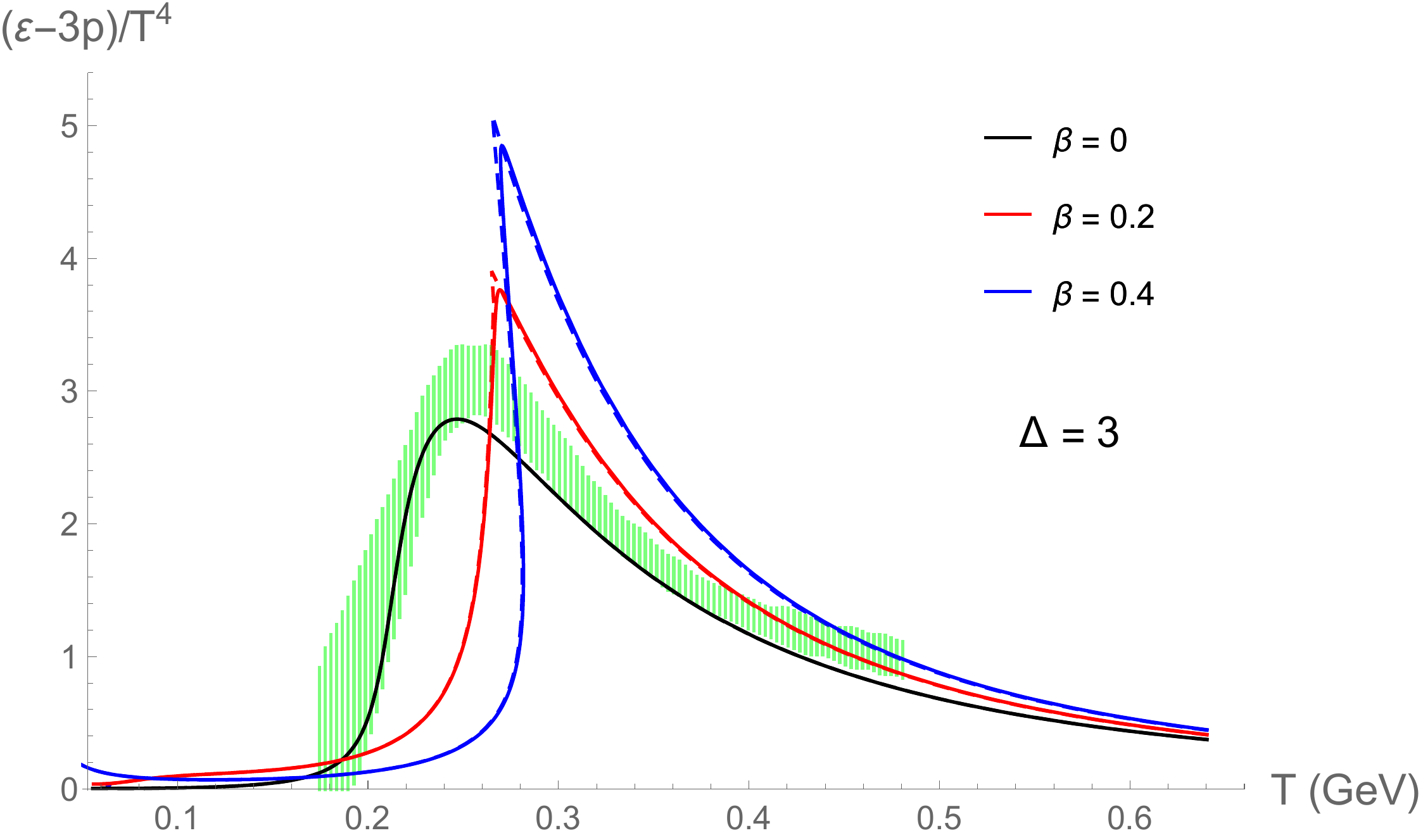} \vskip 0.3cm 
    \end{center}
    \caption{The rescaled energy density $\epsilon/T^4$ (upper panel) and trace anomaly $(\epsilon-3p)/T^4$ (lower panel) for $\beta=0, 0.2, 0.4$ in the case of $\Delta=3$, which are compared with the lattice results of two-flavor QCD represented by the green bands \cite{Burger:2014xga}. The dashed curves denote the case of $m_q=0$ and the solid ones denote the case of $m_q=5\MeV$.}
    \label{fig-e-e3p-T3}
\end{figure}

\begin{figure}
    \centering
    \includegraphics[width=75mm,clip=true,keepaspectratio=true]{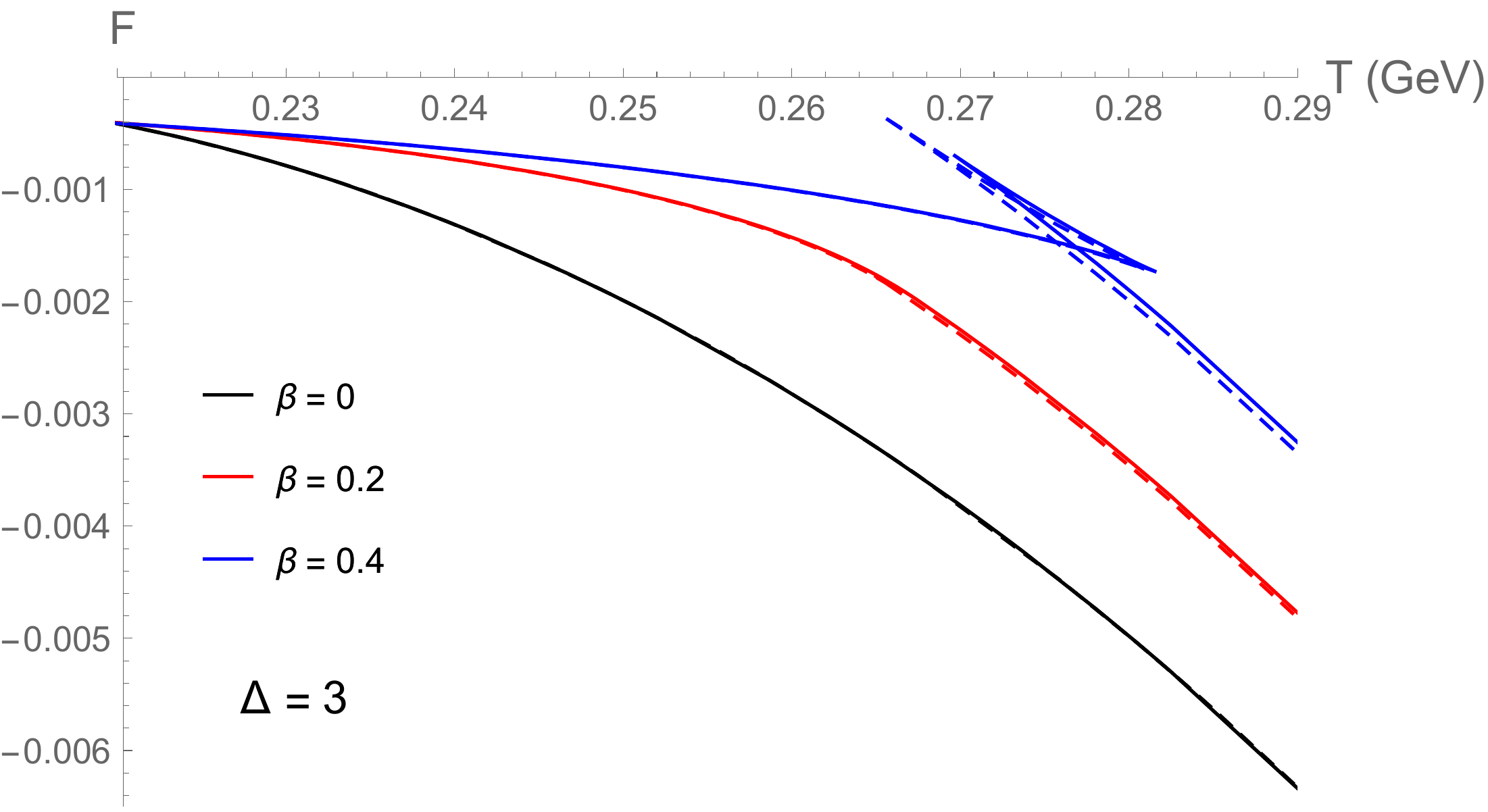}
    \vskip 0.3cm
    \caption{The free energy $F$ as a function of the temperature $T$ for $\beta=0, 0.2, 0.4$ in the case of $\Delta=3$. The dashed and solid curves denote the cases of $m_q=0$ and $m_q=5\MeV$ separately.}
    \label{fig-F-T3}
\end{figure}

The rescaled entropy density $s/T^3$ and pressure $3p/T^4$ as functions of the temperature $T$ are presented in Fig. \ref{fig-s-p-T3}, and the rescaled energy density $\epsilon/T^4$ and trace anomaly $(\epsilon-3p)/T^4$ are presented in Fig. \ref{fig-e-e3p-T3}, where the case of $m_q=0$ has been denoted by the dashed curves which almost coincide with the solid ones of the case of $m_q=5\MeV$. We can see that the equation of state obtained from the model exhibits a crossover behavior in the decoupling case of $\beta=0$ with $m_q=5\MeV$, which mimicks the lattice results of two-flavor QCD quite well. While for the case of $\beta=0.4$, the behaviors of the equation of state indicate a first-order phase transition, which can be seen clearly from the swallow-tailed structure of the free energy $F=-p$ shown in Fig. \ref{fig-F-T3}. As a result, we cannot expect a strong coupling between the flavor sector and the background sector in our setup in order to match with the crossover transition implied by lattice QCD. This makes reasonable the study in Ref. \cite{Fang:2019lsz} with only the decoupling case of $\beta=0$ being addressed.

\begin{figure}
    \centering
    \includegraphics[width=75mm,clip=true,keepaspectratio=true]{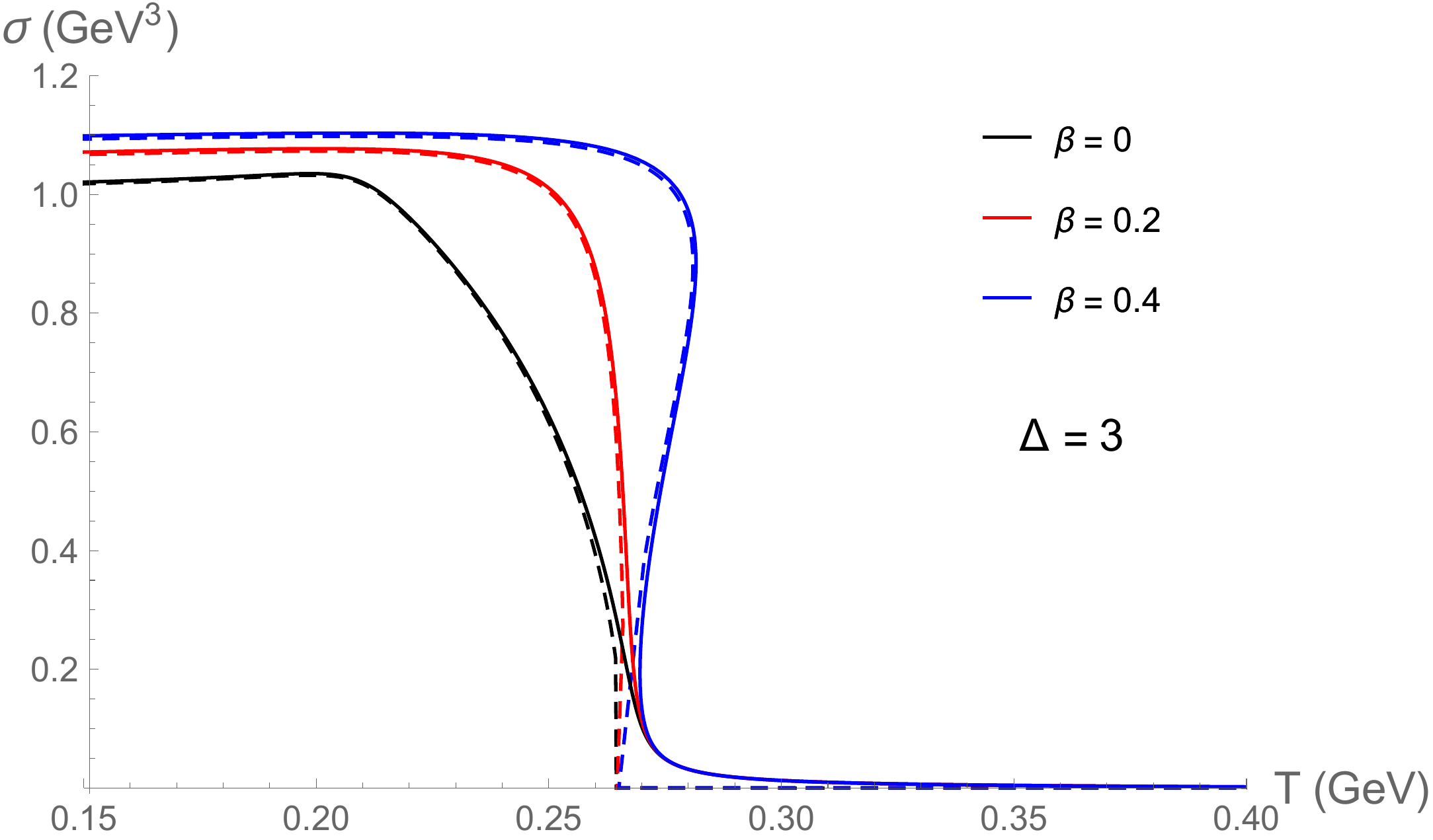}
    \vskip 0.3cm
    \caption{The chiral transition behaviors with respect to the temperature $T$ for $\beta=0, 0.2, 0.4$ in the case of $\Delta=3$. The dashed and solid curves denote the cases of $m_q=0$ and $m_q=5\MeV$ separately.}
    \label{fig-sigma-T3}
\end{figure}

To investigate the properties of chiral transition, we fit the numerical solution of the scalar VEV $\chi(z)$ with the UV asymptotic form (\ref{chi-uv3}) near the boundary $z=0$, so that the chiral condensate $\sigma$ can be extracted for each temperature $T$. The chiral transition behaviors with respect to temperature have been shown in Fig. \ref{fig-sigma-T3}, where we can see that for the decoupling case of $\beta=0$ the chiral transition is a crossover at $m_q=5\MeV$ and becomes a second-order phase transition in the chiral limit with $m_q=0$, which is consistent with the lattice indications \cite{Bhattacharya:2014ara}. With the increase of the coupling constant $\beta$, the chiral transition finally turns into a first-order phase transition, which is exactly the same as that happens in the equation of state, as shown in Fig. \ref{fig-s-p-T3} and Fig. \ref{fig-e-e3p-T3}. Since we address the coupled system of the background fields and the scalar VEV, the behaviors of the equation of state and the chiral transition should be entangled with each other.

\begin{figure}
    \centering
    \includegraphics[width=75mm,clip=true,keepaspectratio=true]{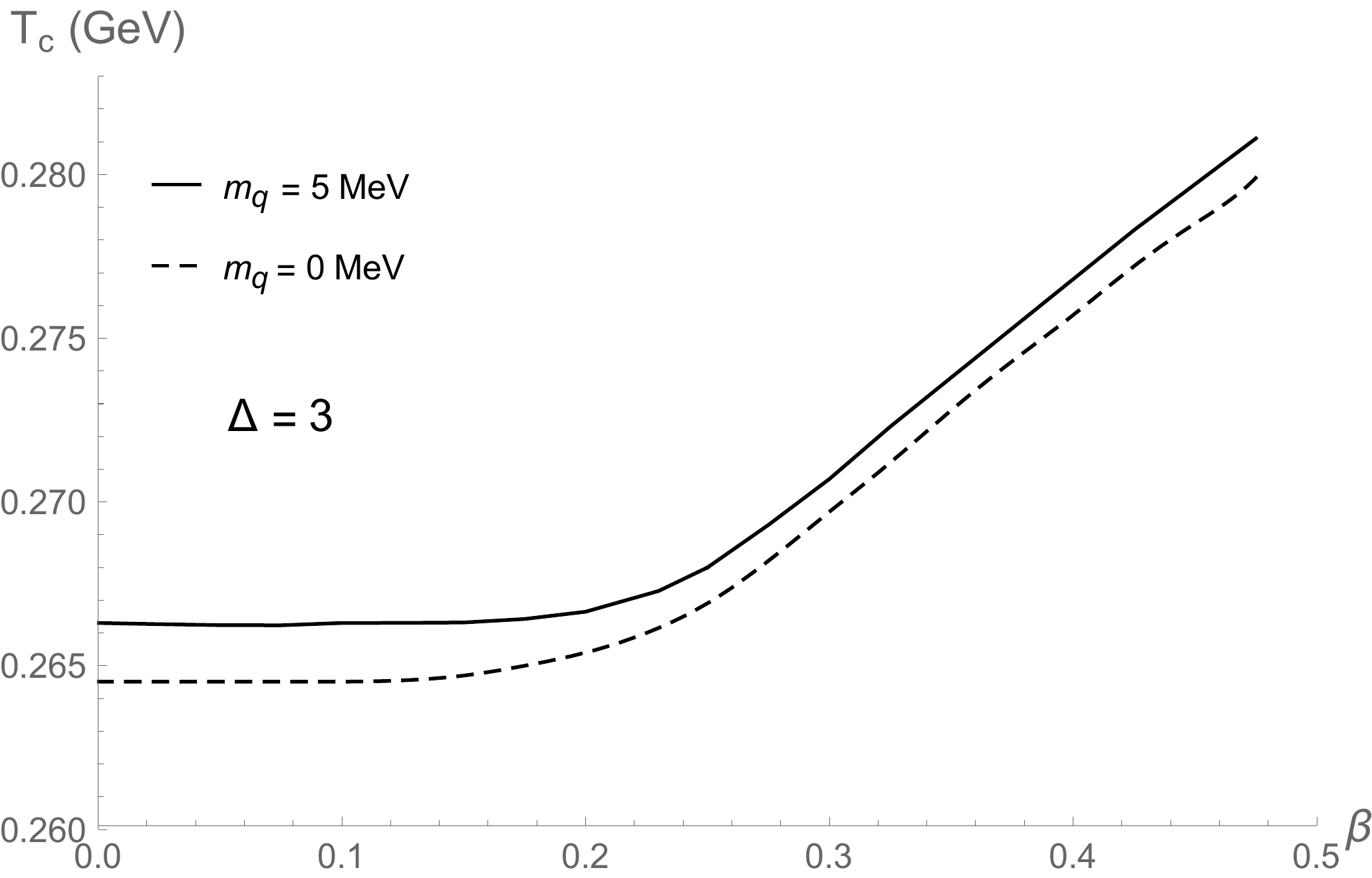}
    \hskip 0.3cm
    Einstein-Dilaton system\caption{The variations of the transition temperature $T_c$ with respect to the coupling constant $\beta$ in the case of $\Delta=3$ with $m_q=0$ and $m_q=5 \MeV$.}
    \label{fig-Tc-beta3}
\end{figure}

We would like to consider the effect of the coupling constant $\beta$ on the transition temperature $T_c$ which may be defined as the extremum point of the curve of $\partial\sigma/\partial T$ for the crossover case with smaller values of $\beta$, while for the case of first-order transitions with larger values of $\beta$ the transition temperature $T_c$ can be easily read from the free energy $F$ as a function of $T$, which has been shown in Fig. \ref{fig-F-T3} for the cases of $\beta=0, 0.2, 0.4$. The dependence of $T_c$ on the coupling constant $\beta$ is shown in Fig. \ref{fig-Tc-beta3}, where we can see that in the decoupling case of $\beta=0$ the transition temperature $T_c\simeq 264.5\MeV$ at $m_q=0$ and $T_c\simeq 266.3\MeV$ at $m_q=5\MeV$. We also find that $T_c$ is almost invariant in the range of $\beta\simeq (0,0.2)$, and then it begins to rise linearly with the increase of $\beta$. The distinctions of $T_c$ as a function of $\beta$ are very small for the cases of $m_q=0$ and $m_q=5\MeV$. We should remark that the transition temperature $T_c$ defined in the crossover case is only apt for chiral transitions, and we can also introduce another $T_c$ which is defined as the extremum point of the first derivative of the equation of state with respect to temperature, which is indeed smaller than the chiral transition temperature in our case. However, this difference will disppear when $\beta$ increases beyond some point such that the transition turns into a first-order one.

\subsection{$\Delta=2.5$ and $3.5$}

We have shown that the QCD equation of state and chiral transition in the two-flavor case can be properly described by the Einstein-dilaton-scalar system with $\Delta=3$ and a smaller coupling between the background and matters. An important issue is how does the scaling dimension $\Delta$ affect the thermodynamic properties in this coupled system. Previous studies indicate that many aspects of thermal QCD can be characterized by the Einstein-dilaton system with some value of $\Delta$ \cite{Noronha:2009ud,Finazzo:2013efa,Finazzo:2014zga}. If the scaling dimension has a significant influence on the equation of state and chiral transition, we may use this effect to determine the proper values of $\Delta$, which actually cannot be done only by theoretical analysis. Therefore, we also consider another two cases with $\Delta=2.5$ and $3.5$ in the BF bound with the aim to investigate the effects of the scaling dimension $\Delta$ on thermodynamics in the Einstein-dilaton-scalar system.

\begin{figure}
    \begin{center}
        \includegraphics[width=68mm,clip=true,keepaspectratio=true]{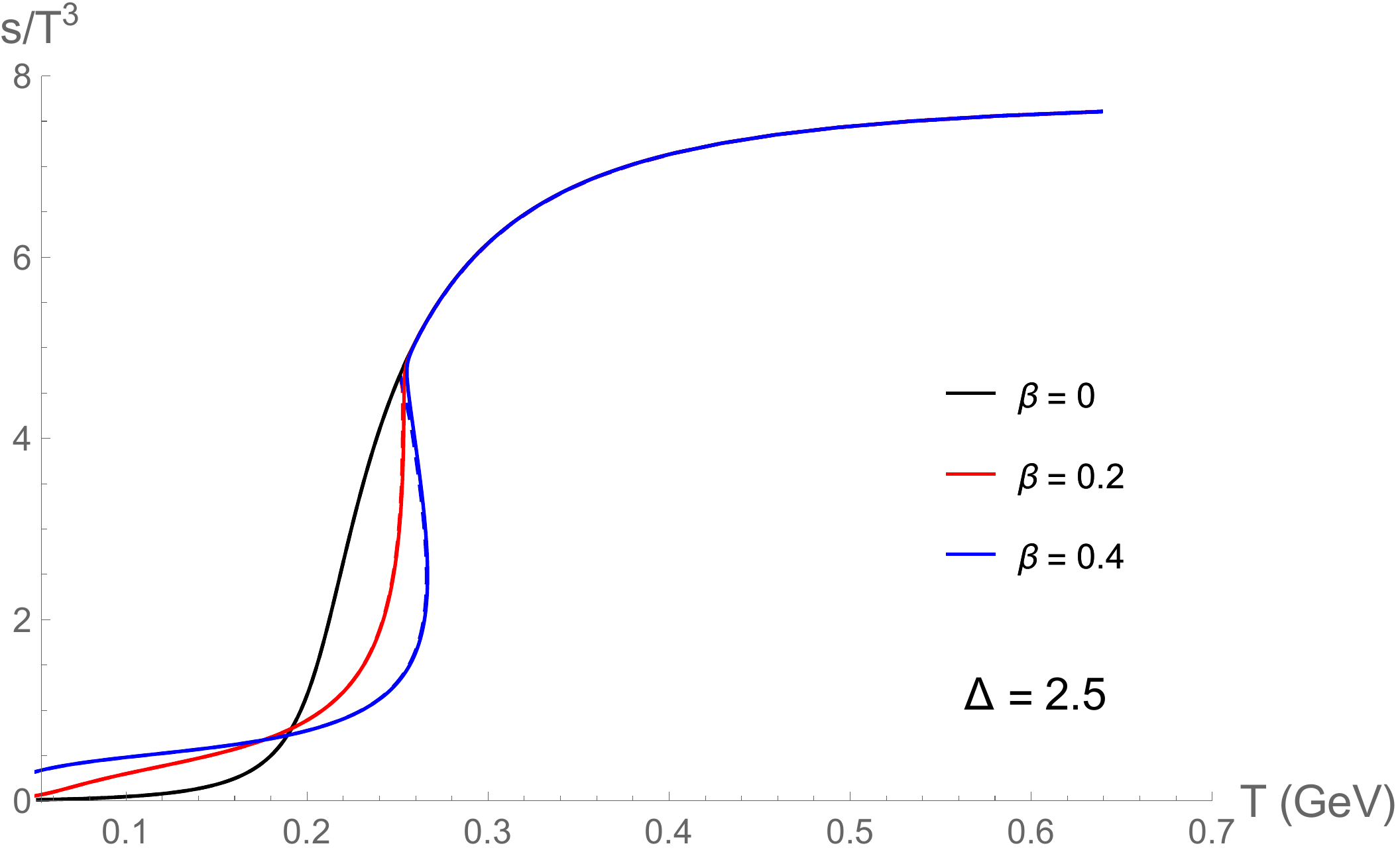}
        \vskip 0.5cm
        \includegraphics[width=68mm,clip=true,keepaspectratio=true]{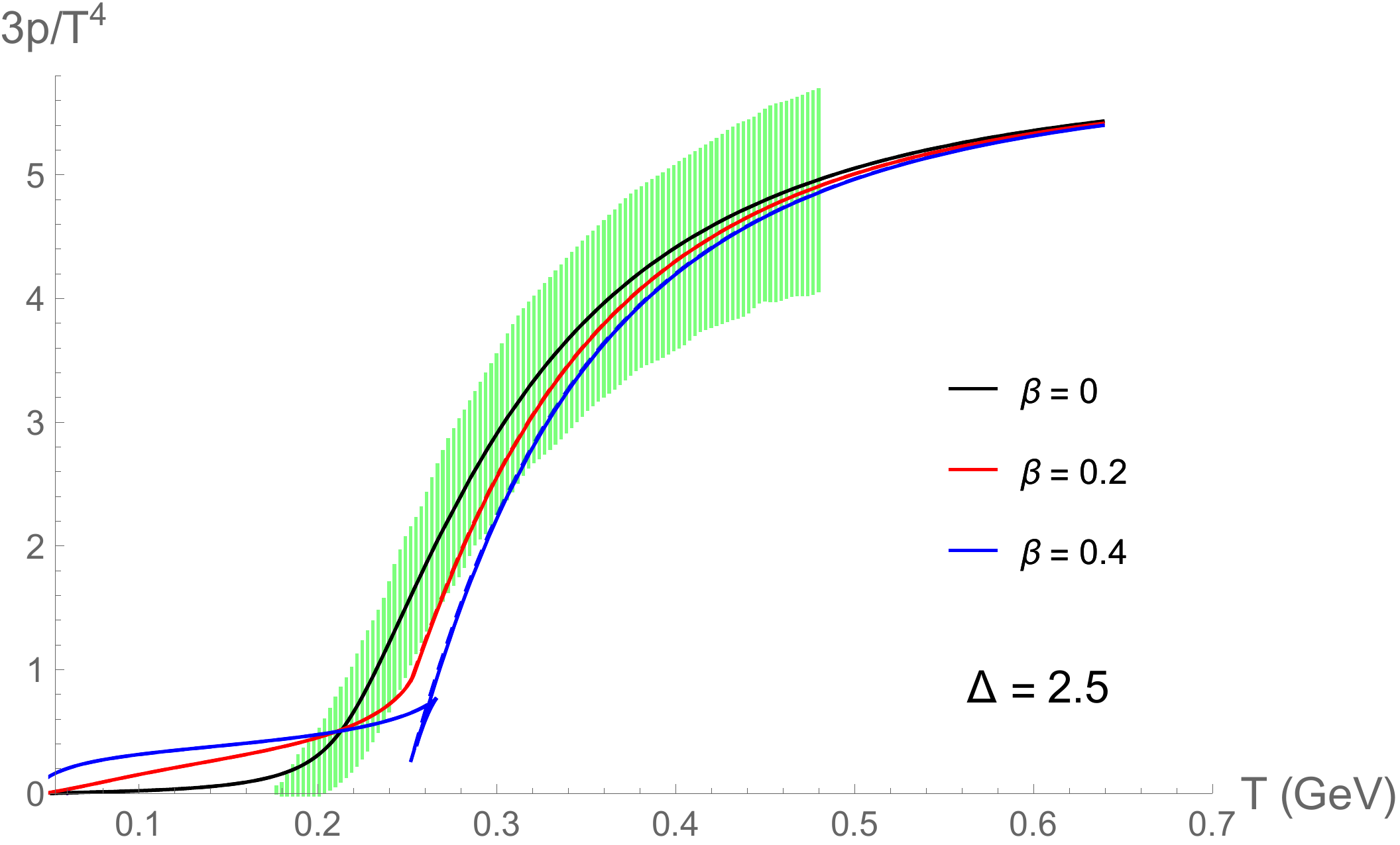} \vskip 0.3cm 
    \end{center}
    \caption{The rescaled entropy density $s/T^3$ (upper panel) and pressure $3p/T^4$ (lower panel) for $\beta=0, 0.2, 0.4$ in the case of $\Delta=2.5$. The green bands represent the lattice interpolations of two-flavor QCD \cite{Burger:2014xga}. The dashed curves denote the case of $m_q=0$ and the solid ones denote the case of $m_q=5\MeV$.}
    \label{fig-s-p-T25}
\end{figure}

\begin{figure}
    \begin{center}
        \includegraphics[width=68mm,clip=true,keepaspectratio=true]{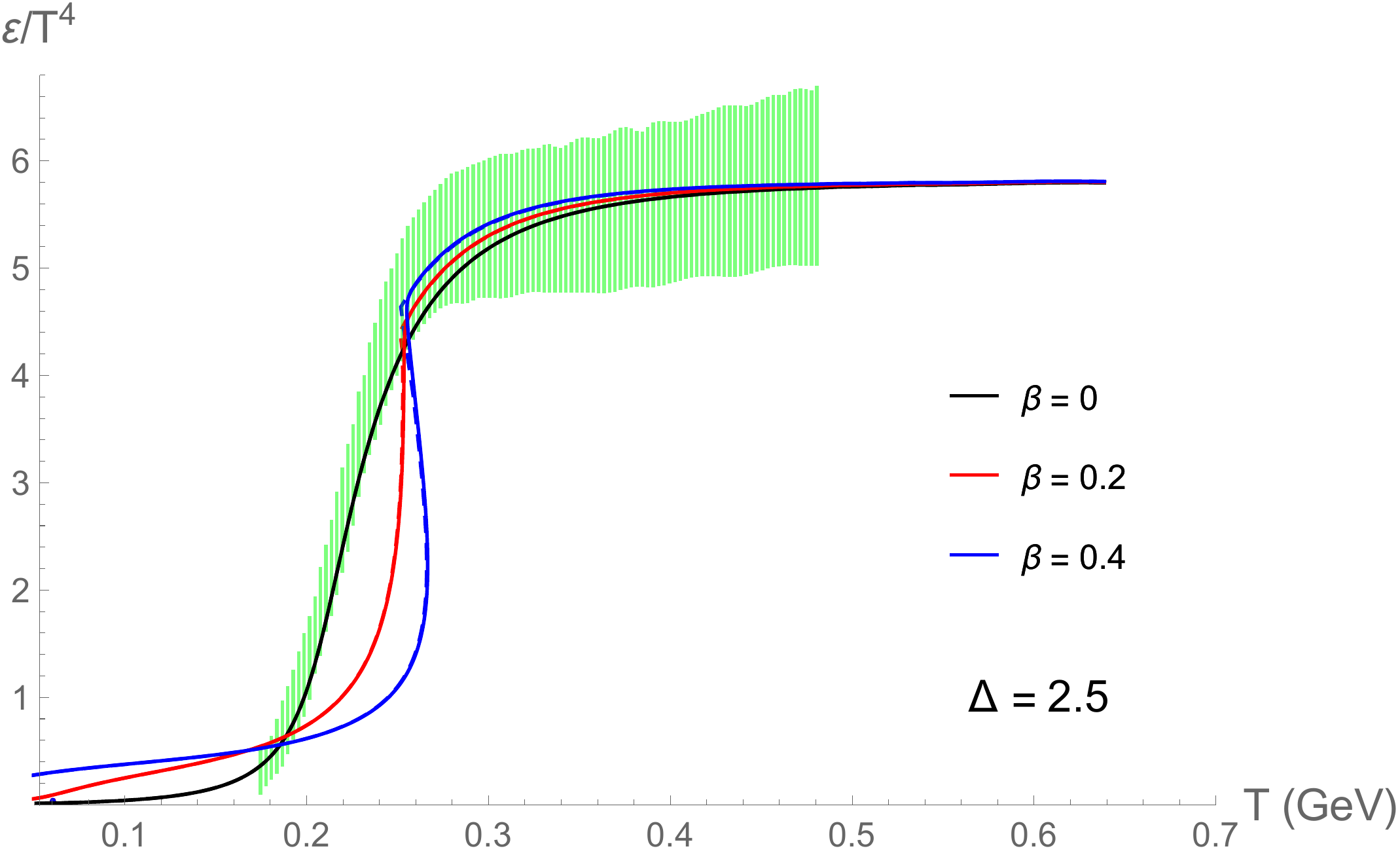}
        \vskip 0.5cm\hspace*{-0.5cm}
        \includegraphics[width=68mm,clip=true,keepaspectratio=true]{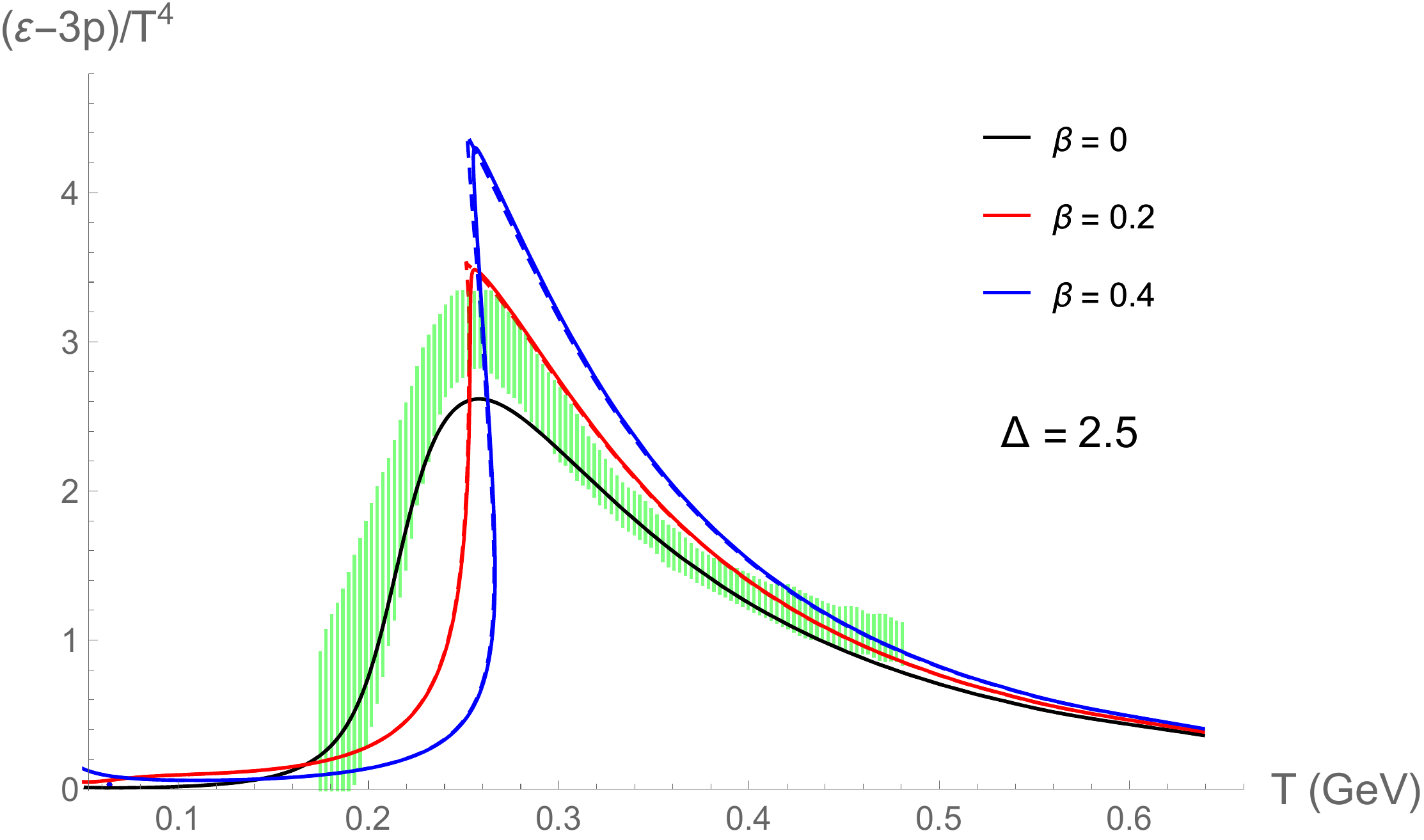} \vskip 0.3cm 
    \end{center}
    \caption{The rescaled energy density $\epsilon/T^4$ (upper panel) and trace anomaly $(\epsilon-3p)/T^4$ (lower panel) for $\beta=0, 0.2, 0.4$ in the case of $\Delta=2.5$, which are compared with the lattice simulations of two-flavor QCD represented by the green bands \cite{Burger:2014xga}. The dashed curves denote the case of $m_q=0$ and the solid ones denote the case of $m_q=5\MeV$.}
    \label{fig-e-e3p-T25}
\end{figure}

\begin{figure}
    \centering
    \includegraphics[width=75mm,clip=true,keepaspectratio=true]{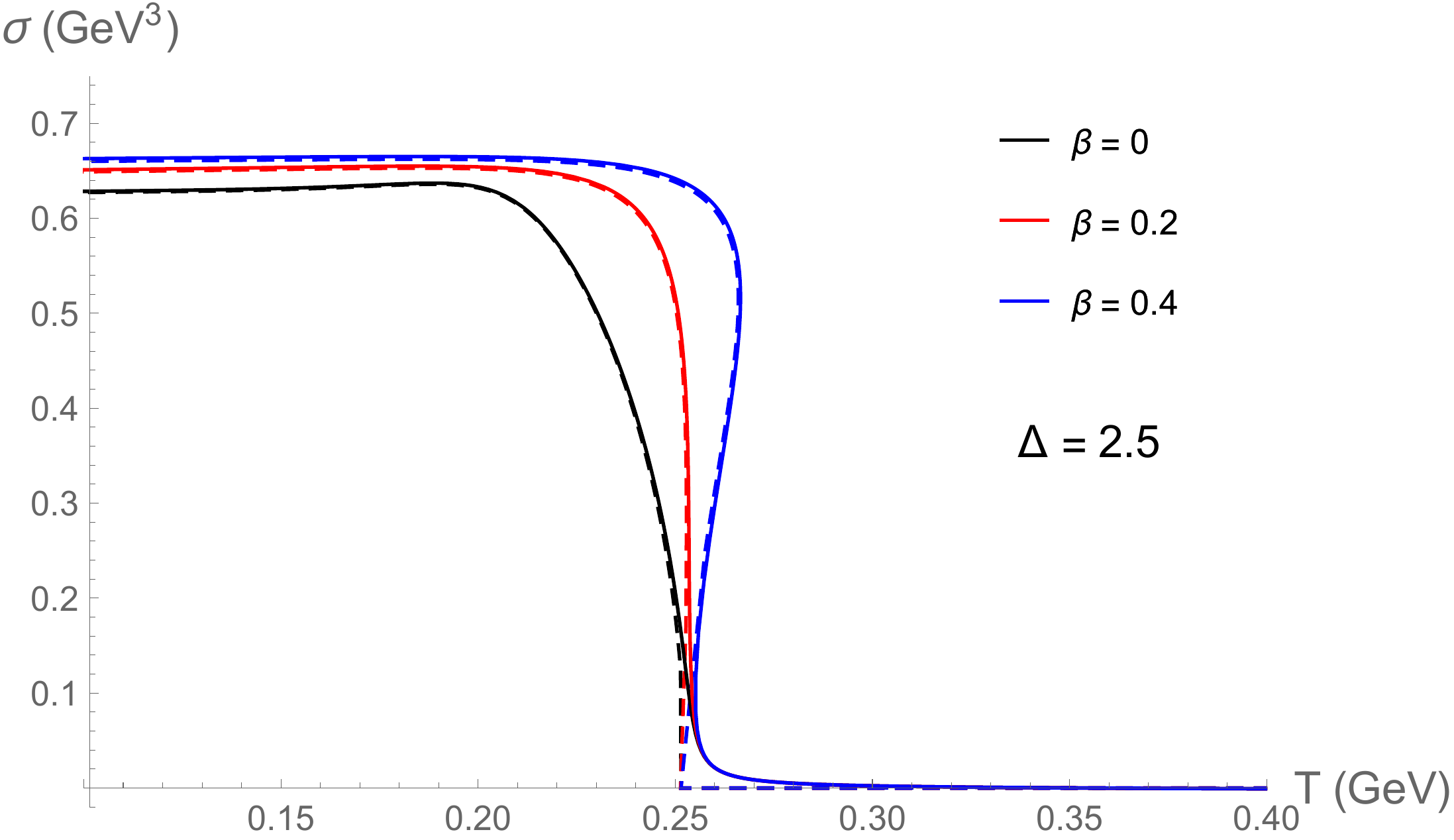}
    \vskip 0.3cm
    \caption{The chiral transition behaviors with respect to the temperature $T$ for $\beta=0, 0.2, 0.4$ in the case of $\Delta=2.5$. The dashed and solid curves denote the cases of $m_q=0$ and $m_q=5\MeV$ separately.}
    \label{fig-sigma-T25}
\end{figure}

We first consider the case of $\Delta=2.5$ and compute the equation of state that will be matched with the lattice results of two-flavor QCD for the decoupling case of $\beta=0$ with $m_q=5\MeV$. The fitting parameters are taken as $\gamma=0.5$, $b_4=-0.125$ and $p_1=0.95\GeV$. The model results of the rescaled entropy density $s/T^3$ and pressure $3p/T^4$ are shown in Fig. \ref{fig-s-p-T25}, and the rescaled energy density $\epsilon/T^4$ and trace anomaly $(\epsilon-3p)/T^4$ are shown in Fig. \ref{fig-e-e3p-T25}. It can be seen that the equation of state obtained in the decoupling case can also fit the lattice results well, and the crossover transition changes into a first-order one with the increase of the coupling constant $\beta$ in the same manner as that in the case of $\Delta=3$. Thus it seems impossible to distinguish different values of $\Delta$ only through the equation of state, as indicated in the previous studies \cite{Gubser:2008yx,Noronha:2009ud}. This is one of the reasons why we resort to considering the Einstein-Dilaton-scalar system, which allows us to investigate both the equation of state and the chiral transition. We show the chiral transition behaviors for the case of $\Delta=2.5$ in Fig. \ref{fig-sigma-T25}, where we find that they have the similar dependence on the coupling constant $\beta$ as that in the case of $\Delta=3$. However, the chiral transition temperature and also the absolute value of the chiral condensate become smaller in this case, as compared to the case of $\Delta=3$.

\begin{figure}
    \begin{center}
        \includegraphics[width=68mm,clip=true,keepaspectratio=true]{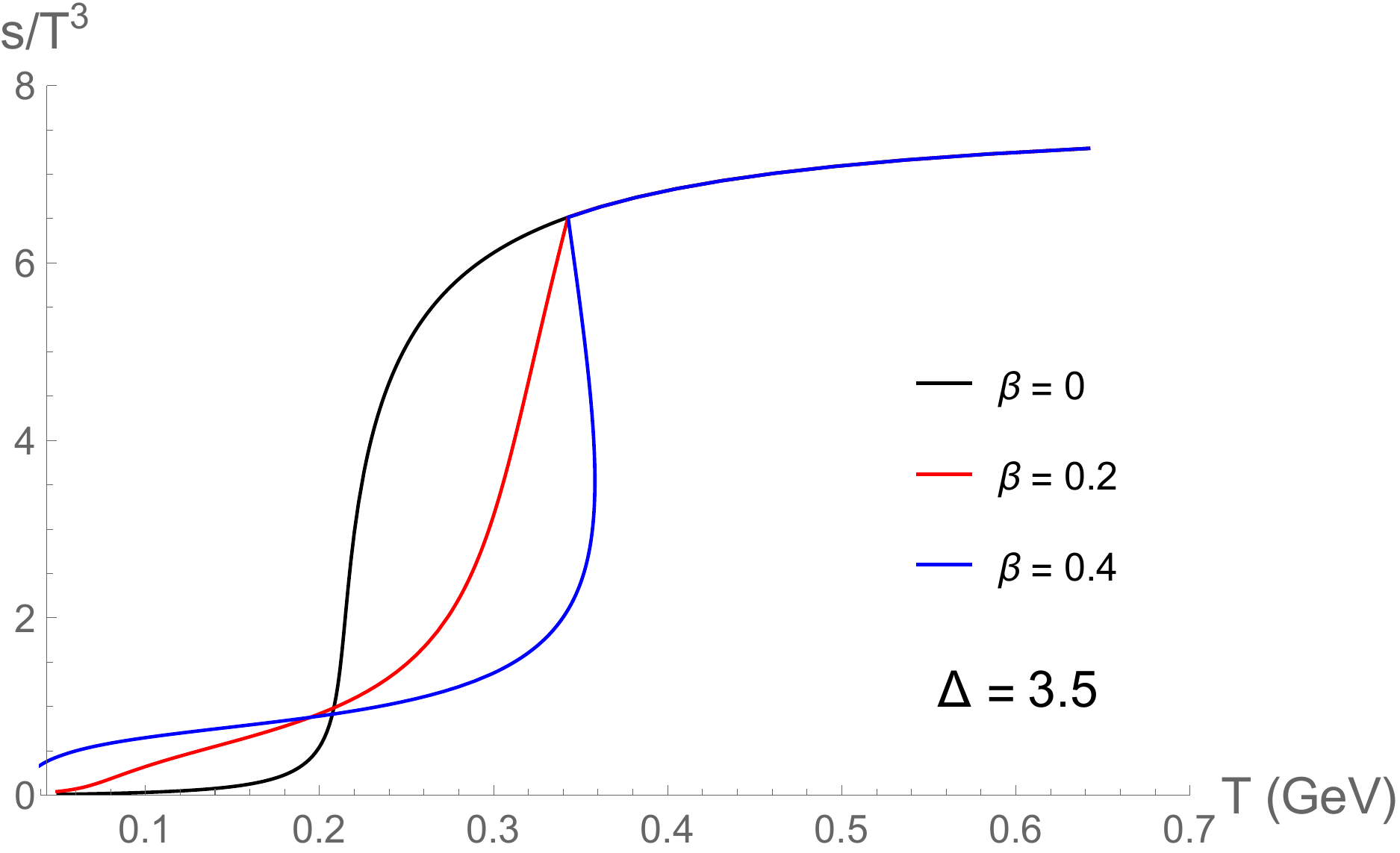}
        \vskip 0.5cm
        \includegraphics[width=68mm,clip=true,keepaspectratio=true]{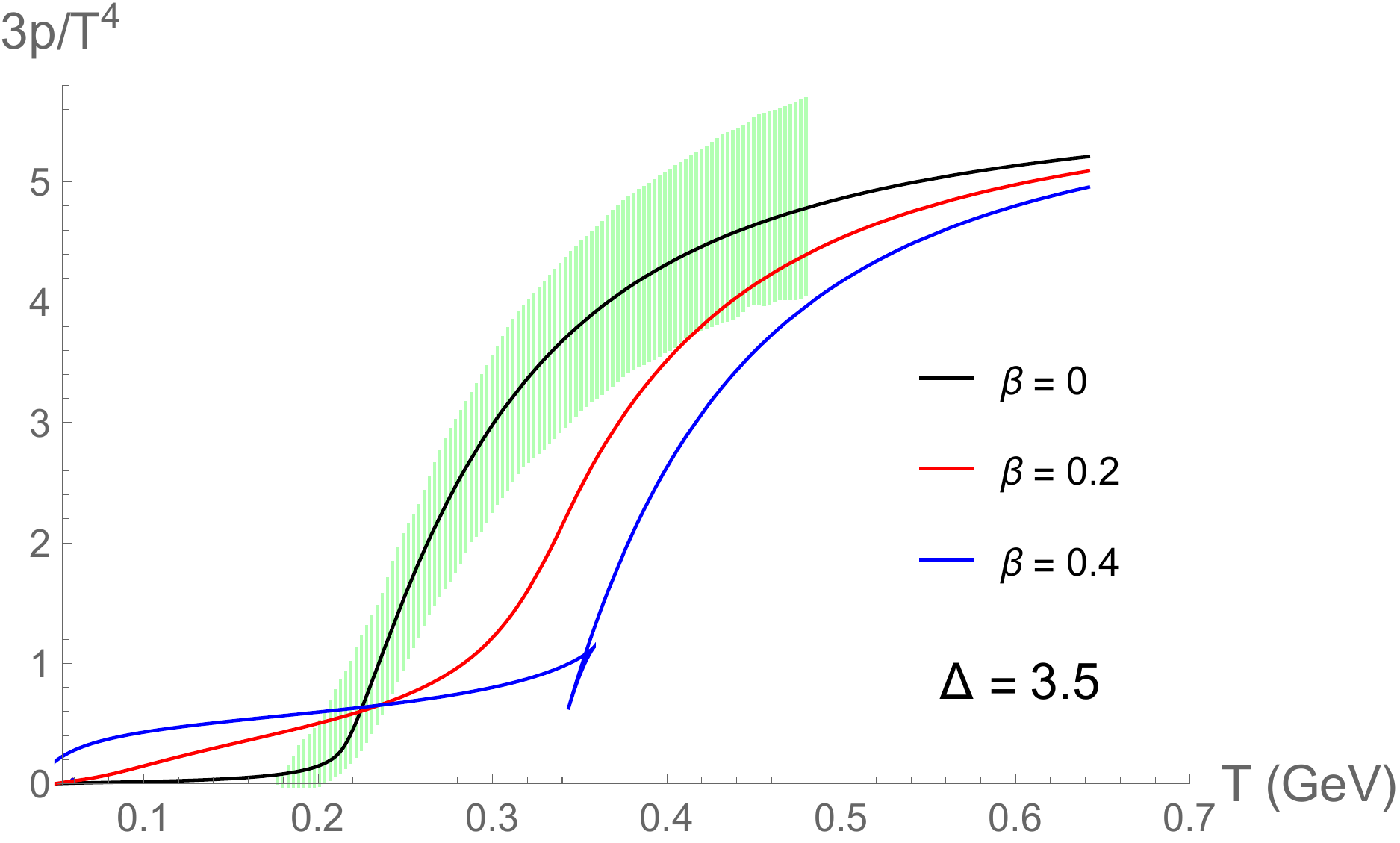} \vskip 0.3cm 
    \end{center}
    \caption{The rescaled entropy density $s/T^3$ (upper panel) and pressure $3p/T^4$ (lower panel) for $\beta=0, 0.2, 0.4$ in the case of $\Delta=3.5$ with $m_q=0$. The green bands represent the lattice interpolations of two-flavor QCD \cite{Burger:2014xga}.}
    \label{fig-s-p-T35}
\end{figure}

\begin{figure}
    \begin{center}
        \includegraphics[width=68mm,clip=true,keepaspectratio=true]{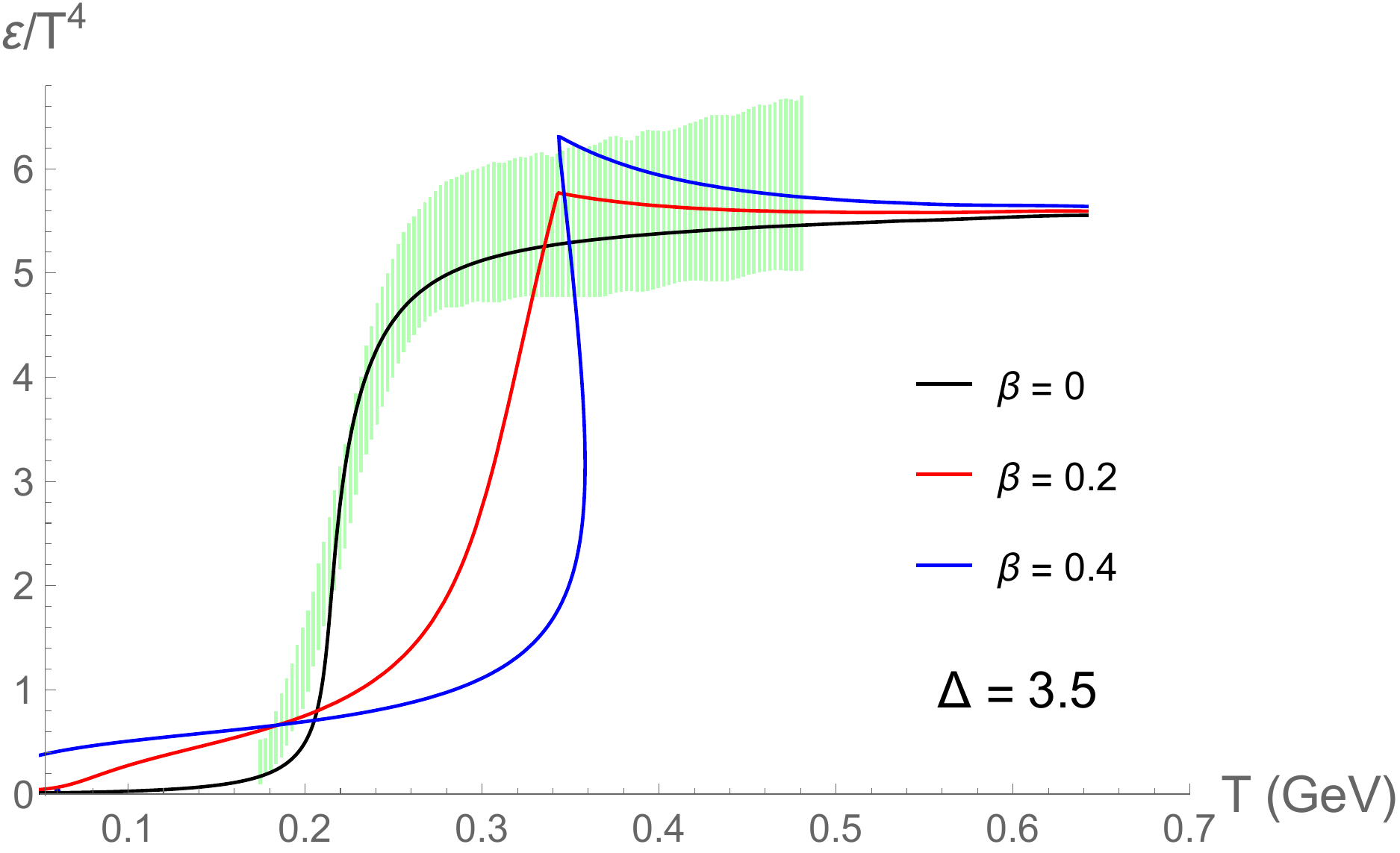}
        \vskip 0.5cm\hspace*{-0.5cm}
        \includegraphics[width=68mm,clip=true,keepaspectratio=true]{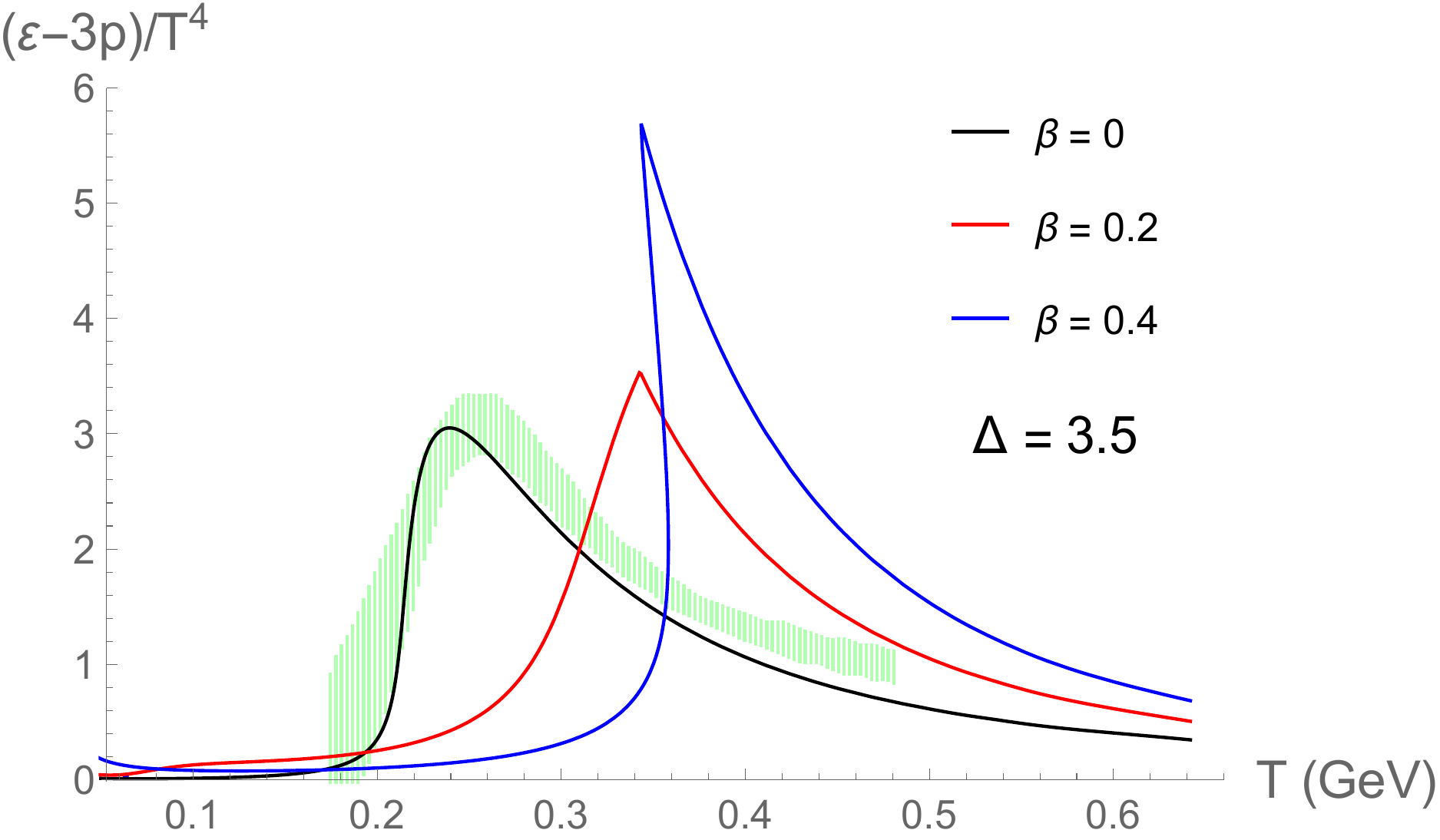} \vskip 0.3cm 
    \end{center}
    \caption{The rescaled energy density $\epsilon/T^4$ (upper panel) and trace anomaly $(\epsilon-3p)/T^4$ (lower panel) for $\beta=0, 0.2, 0.4$ in the case of $\Delta=3.5$ with $m_q=0$, which are compared with the lattice results of two-flavor QCD represented by the green bands \cite{Burger:2014xga}.}
    \label{fig-e-e3p-T35}
\end{figure}

\begin{figure}
    \centering
    \includegraphics[width=75mm,clip=true,keepaspectratio=true]{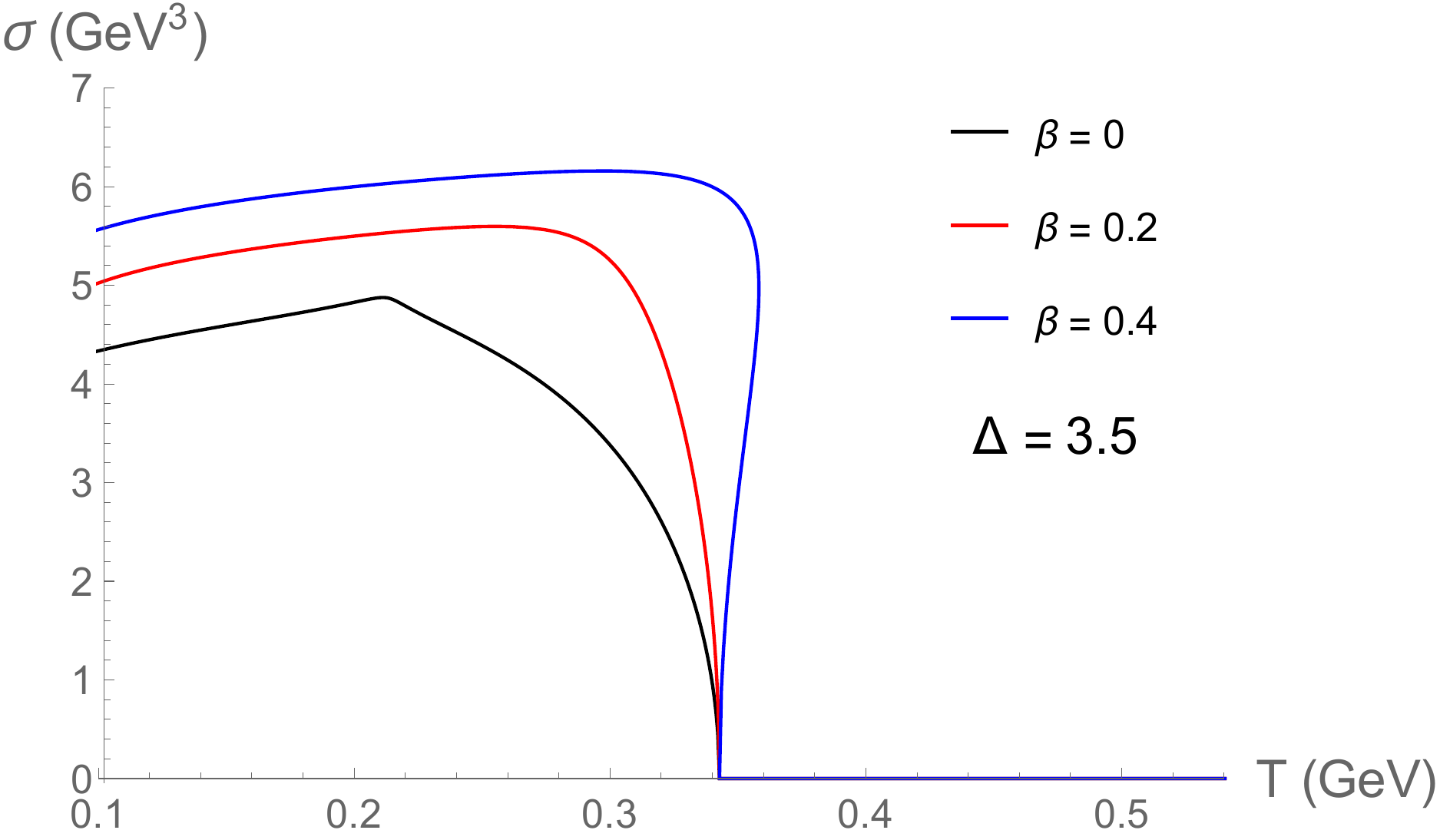}
    \vskip 0.3cm
    \caption{The chiral transition behaviors with respect to the temperature $T$ for $\beta=0, 0.2, 0.4$ in the case of $\Delta=3.5$ with $m_q=0$.}
    \label{fig-sigma-T35}
\end{figure}

We then repeat the computation for the equation of state and chiral transition in the case of $\Delta=3.5$, but only restrict to the chiral limit with $m_q=0$. The model parameters are set to $\gamma=0.2$, $b_4=-0.175$ and $p_1=0.6\GeV$ in order to match with the two-flavor lattice results in the decoupling case of $\beta=0$. Note that the quark mass does not affect the equation of state for $\beta=0$. We present the model results of the rescaled entropy density $s/T^3$ and pressure $3p/T^4$ in Fig. \ref{fig-s-p-T35} and the rescaled energy density $\epsilon/T^4$ and trace anomaly $(\epsilon-3p)/T^4$ in Fig. \ref{fig-e-e3p-T35}. We also show in Fig. \ref{fig-sigma-T35} the behaviors of chiral transition which changes from a second-order phase transition into a first-order one with the increase of $\beta$, just as in the cases of $\Delta=2.5$ and $\Delta=3$ with $m_q=0$. An apparent difference is that the chiral transition temperature in the case of $\Delta=3.5$ is much larger than those in the former cases. Moreover, we also find that the influence of the coupling constant $\beta$ on the equation of state and chiral transition becomes more and more significant with the increase of the scaling dimension $\Delta$, which cannot be shown in the Einstein-dilaton system or in the decoupling case of $\beta=0$ \cite{Fang:2019lsz}.

\begin{figure}
    \centering
    \includegraphics[width=75mm,clip=true,keepaspectratio=true]{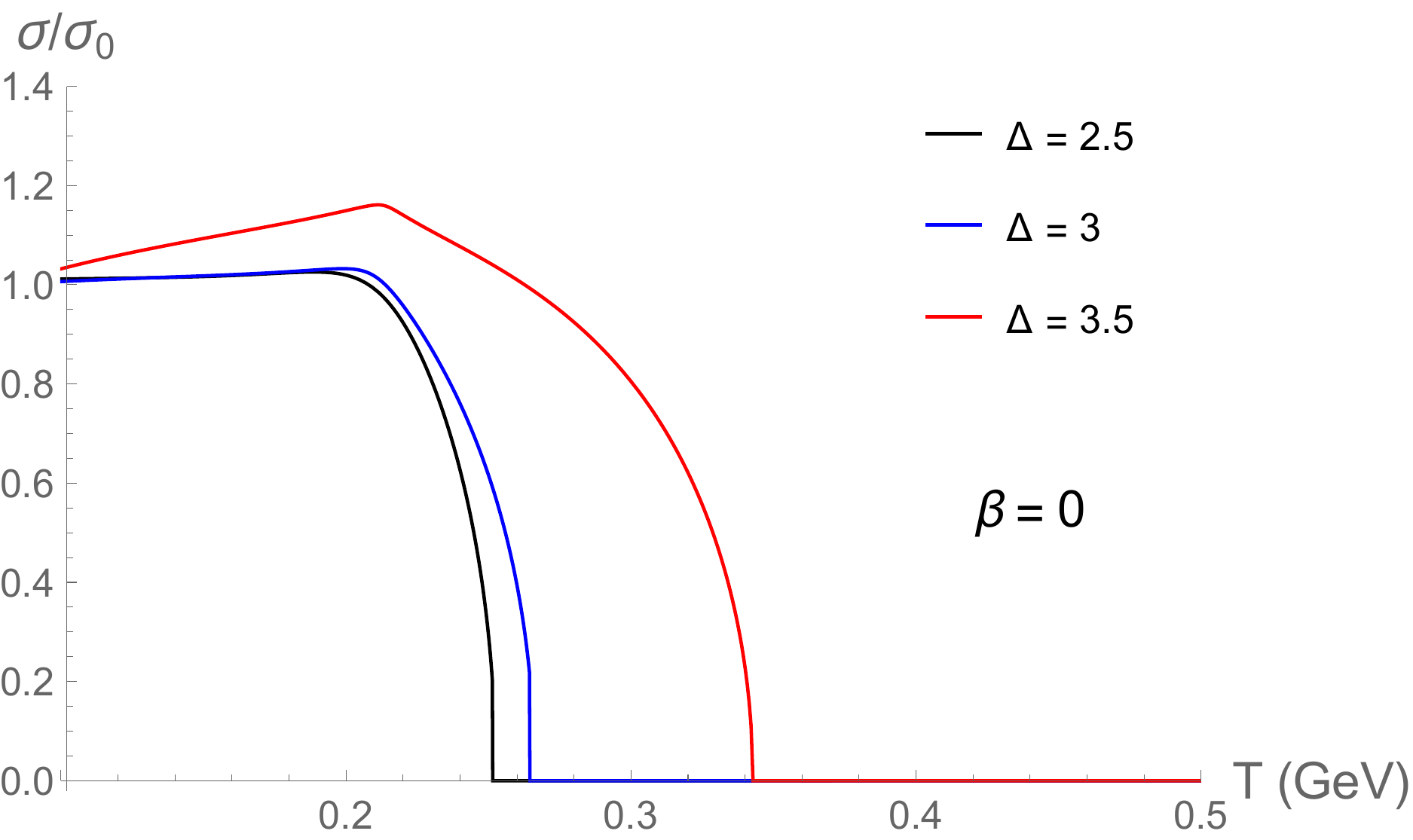}
    \vskip 0.3cm
    \caption{The behaviors of the rescaled chiral condensate with respect to the temperature $T$ for $\Delta=2.5, 3$ and $3.5$ in the decoupling case of $\beta=0$ with $m_q=0$.}
    \label{fig-3sigma-T}
\end{figure}

As we have seen, when the background of the coupled system is fixed by the QCD equation of state, the effects of the scaling dimension $\Delta$ are manifested in chiral transition. In Fig. \ref{fig-3sigma-T}, we show for clarity the chiral transition behaviors in terms of the rescaled chiral condensate for $\Delta=2.5, 3, 3.5$ in the decoupling case of $\beta=0$ with zero quark mass, from which we can see obviously that the chiral transition temperature increases with the increase of the scaling dimension $\Delta$. Hence, in contrast to the Einstein-dilaton system, the Einstein-dilaton-scalar system can be used to distinguish different values of $\Delta$ that correspond to the dimensions of the dual operator of the dilaton at different energy scales. This is sensible, considering that the flavor sector of the coupled system characterizes the low-energy hadron physics which must be related to some specific energy scale like the chiral scale, and the chiral dynamics should come in to select such a scale that plays a significant role in the holographic framework of the Einstein-dilaton-scalar system.

\section{Conclusion and discussion}\label{sec-conc}

In this work, we consider an improved soft-wall AdS/QCD model coupled to an Einstein-dilaton system which can be seen as a dual of the boundary QCD with both the pure Yang-Mills sector and the flavor sector. The correlation between the deconfining and chiral transitions was investigated in detail in the so-called Einstein-dilaton-scalar system with the bulk scalar field representing the vacuum of matters in the improved soft-wall model. There have been many researches on the interrelation between these two kinds of QCD transitions, and our work provides a preliminary attempt to address this issue in the framework of bottom-up AdS/QCD.

The equation of state and the chiral transition have been studied for the cases of $\Delta=2.5, 3, 3.5$, and in each case we take three values of the coupling constant $\beta$ for computation, that is, $\beta=0, 0.2, 0.4$. We find that for each value of $\Delta$ the equation of state can be well matched with the lattice results of two-flavor QCD in the decoupling case of $\beta=0$, which implies that the scaling dimension $\Delta$ is not unique for the description of the properties of deconfinement in the Einstein-dilaton system. Essentially, this is due to the redundant degrees of freedom in the dilaton potential which cannot be determined from the first principle. As a phenomenological model, we then resort to other properties of QCD phase transition in order to handle this issue. We consider the Einstein-dilaton system integrated with the soft-wall AdS/QCD model, which allows us to address the deconfining and chiral transitions simultaneously. We find that these two transitions are tightly correlated with each other under the influence of the coupling constant $\beta$ in this coupled system of background and matters.

In contrast to that of the Einstein-dilaton system, the scaling dimension $\Delta$ plays a significant role in the description of QCD phase transition in the Einstein-dilaton-scalar system. We find that the value of $\Delta$ has a prominent effect on the behaviors of the deconfining and chiral transitions, especially in the situation with nonzero coupling constant $\beta$. Although the equation of state can all be matched with the lattice results in the decoupling case of $\beta=0$, the chiral transition behaviors show distinctions for different values of $\Delta$, and particularly the transition temperature $T_c$ increases with the increase of $\Delta$, which is more obvious for larger values of $\Delta$ in the BF bound, as can be seen in Fig. \ref{fig-3sigma-T}. This is a merit of the Einstein-dilaton-scalar system which offers a way to specify the scaling dimension of the dual operator of the dilation from phenomenology. As we know, the scalar VEV embodies the informations of the low-energy hadron physics and thus sets a special energy scale at which the scaling dimension $\Delta$ should be computed.

In the weak-coupling case with nonzero quark mass, both the equation of state and the chiral transition exhibit a crossover behavior and turn into first-order phase transition with the increase of $\beta$. Hence, the coupling between the background fields and the scalar VEV cannot be strong in order to match with the lattice results of two-flavor QCD. In other words, the back-reaction of the flavor sector to the background should be as small as possible, which supports the previous studies of AdS/QCD based on a fixed bulk background with no back-reaction effects. One characteristic of the Einstein-dilaton-scalar system is that the chiral transition temperature $T_{\chi}$ is higher than the deconfinement temperature $T_{d}$ implied by the equation of state, and the discrepancy between these two transition temperatures becomes larger and larger with the increase of $\Delta$. There are still many debates on the relation between $T_{\chi}$ and $T_{d}$ \cite{Suganuma:2017syi}. General arguments from bag models support that $T_{\chi}>T_{d}$ \cite{Karsch:2001cy}, while lattice QCD seems to imply the inverse result \cite{Burger:2014xga}. On the other hand, there are also lattice simulations indicating that these two transition temperatures are very close to each other \cite{Ding:2015ona}, which, though, does not exclude the possibility to separate the scales of chiral symmetry breaking and confinement \cite{Evans:2020ztq}.

Many issues need to be clarified in the further study. We shall proceed to investigate the QCD phase diagram at finite chemical potential in the Einstein-dilaton-scalar system. We may also need to consider other forms of the dilaton potential in order to reproduce the realistic phase structure of QCD. As we know, the pure Yang-Mills theory admits a first-order phase transition, while for QCD with physical quark masses this is more likely a crossover transition. Moreover, how to realize the linear confinement and to give a consistent description for hadron spectra is still an inconclusive issue in this holographic framework.

\section*{Acknowledgements}
This work is supported by the National Natural Science Foundation of China (NSFC) under Grant No. 11905055, the Natural Science Foundation of Hunan Province, China under Grant No. 2020JJ5026 and the Fundamental Research Funds for the Central Universities.

\bibliography{refs-AdSQCD}

\end{document}